\documentclass{article}
\usepackage[utf8]{inputenc}
\usepackage{authblk}
\usepackage[colorlinks]{hyperref}
\usepackage[normalem]{ulem}

\title{On a Generating Function for the Isotropic Basis Functions and Other Connected Results}

\author[1,2*]{Zachary Slepian}
\author[1]{Jessica Chellino}
\author[1,3]{Jiamin Hou}
\author[1]{Alessandro Greco}
\affil[1]{\footnotesize Department of Astronomy, University of Florida, 211 Bryant Space Science Center, Gainesville, FL 32611, USA}
\affil[2]{\footnotesize Physics Division, Lawrence Berkeley National Laboratory, 1 Cyclotron Road, Berkeley, CA 94709, USA}
\affil[3]{Max-Planck-Institut f\"ur extraterrestrische Physik, Gießenbachstraße 1, 85748 Garching, Germany}
\affil[*]{Electronic Address: \href{mailto: zslepian@ufl.edu}{zslepian@ufl.edu}}

\usepackage[a4paper, total={6in, 8in}]{geometry}
\usepackage{amsmath}
\usepackage{amssymb}

\usepackage[numbers]{natbib}
\usepackage{graphicx}
\usepackage{subcaption}
\renewcommand{\vec}[1]{\boldsymbol{#1}}

\usepackage{color}
\usepackage{enumitem}
\def\beq{\begin{eqnarray}}
\def\eeq{\end{eqnarray}}

\usepackage[dvipsnames]{xcolor}

\definecolor{darkgreen}{RGB}{0,120,0}

\begin{document}

\maketitle

\begin{abstract}
Recently isotropic basis functions of $N$ unit vector arguments were presented; these are of significant use in measuring the N-Point Correlation Functions (NPCFs) of galaxy clustering. Here we develop the generating function for these basis functions---\textit{i.e.} that function which, expanded in a power series, has as its angular part the isotropic functions. We show that this can be developed using basic properties of the plane wave. A main use of the generating function is as an efficient route to obtaining the Cartesian basis expressions for the isotropic functions.  We show that the methods here enable computing difficult overlap integrals of multiple spherical Bessel functions, and we also give related expansions of the Dirac Delta function into the isotropic basis. Finally, we outline how the Cartesian expressions for the isotropic basis functions might be used to enable a faster NPCF algorithm on the CPU.

\end{abstract}

\section{Introduction}
Many situations in physics invoke rotational invariance; particularly, in cosmology, the distribution of galaxies around a given center should appear isotropic, modulo distortions induced by conversion of redshift into position. Hence, functions that are invariant under simultaneous rotation of all of their unit-vector arguments are particularly useful. In \cite{Cahn}, these isotropic basis functions were developed, and their importance and other related previous work is more fully outlined there. The idea of using them to parameterize N-Point Correlation Functions, quantifying the spatial clustering of the distribution of galaxies, was presented in \cite{encore}, with a detection of the even-parity, ``connected''  (due to nonlinear gravitational evolution) 4-Point Correlation Function (4PCF) in \cite{philcox_four}. The idea of using the decomposition of the 4PCF into even and odd-parity 3-argument basis functions to search for parity violation first presented in \cite{parity-prl}, and the covariance matrix required for this search computed in the isotropic basis in \cite{hou-covar}. The first detection of parity violation in 3D large-scale structure was made in \cite{hou-parity}, with $7\sigma$ evidence in one sample and $3\sigma$ evidence in another smaller sample, both of the Sloan Digital Sky Survey Baryon Oscillation Spectroscopic Survey (SDSS BOSS). Another work \cite{philcox-parity} also found evidence for parity violation in CMASS, using the same covariance matrix template.\footnote{This work should unfortunately not be viewed as an independent confirmation, since it was done when its author was a member of our group, using a preliminary covariance matrix template computed by a member of our group. This template was not fully calibrated to the final BOSS data used. This work also uses incorrect systematics weights on the galaxies. This was the result of selecting data from the BOSS combined (LOWZ+CMASS) sample by imposing a redshift cut to include just the CMASS portion, but then using weights appropriate to the full sample, not the subset selected.} This latter work also used the isotropic basis functions to perform calculations of an inflationary model, and subsequent works have also done so, \textit{e.g.} \cite{cabass}. The isotropic basis functions have also been used to compute a model of the even-parity, connected 4PCF (\textit{i.e.} that due to gravity, beyond the Gaussian Random Field contribution) in \cite{william}. Furthermore, they have been used to explore the signature of Baryon Acoustic Oscillations (BAO, \textit{e.g.} \cite{se_simple}) in the odd-parity 4PCF as a means of verifying if it is genuine \cite{bao_odd}.

Here, our goal is to provide generating functions for these isotropic basis functions, as well as present a few useful associated results and applications. 

First, we briefly recall what a generating function is. A frequent example in physics is the multipole expansion, wherein one has
\begin{align}
\label{eqn:multip}
\frac{1}{ | \vec{r}_1 - \vec{r}_2 | } = \sum_{\ell = 0}^{\infty} \frac{r_<^{\ell}}{r_>^{\ell + 1}} \mathcal{L}_{\ell}(\hat{r}_1 \cdot \hat{r}_2).
\end{align}
Here, we have an expansion of the potential between two points at $\vec{r}_1$ and $\vec{r}_2$, and the left-hand side is the ``generating function'' of the Legendre polynomials as the $\ell^{\rm th}$ Legendre polynomial, $\mathcal{L}_{\ell}$, can be picked out if we focus on the right-hand side term proportional to the $\ell^{th}$ power of $r_<$. We may pull out $r_<$ from both sides and work in terms of $t \equiv (r_</r_>) \leq 1$, such that the series' convergence is manifest and by expanding the vector sum magnitude on the left-hand side we may write
\begin{align}
\frac{1}{\sqrt{x^2 - 2 x t + 1}} = \sum_{\ell = 0}^{\infty} t^{\ell} \mathcal{L}_{\ell}(x),
\end{align}
where we have defined $x \equiv \hat{r}_1 \cdot \hat{r}_2 $. The left-hand side above is the most common form for the ``generating function'' of the Legendre polynomials, which themselves form an isotropic basis set for functions of two unit vectors (\textit{i.e.} $\hat{r}_1$ and $\hat{r}_2$; since $x$ is the dot product it is invariant under simultaneous rotation of $\hat{r}_1$ and $\hat{r}_2$). Since the isotropic basis functions of \cite{Cahn} are essentially generalizations of the Legendre polynomials to more than two unit vectors, it seems desirable to have a generating function for them as well. Motivated by work proving the addition theorem for regular solid harmonics using the plane wave expansion \cite{deb, Chak}, we do so here.

This work is structured as follows. First, we present the basic expansion we will use and show how it works straightforwardly for $N = 2$ and $3$ basis functions (\S\ref{sec:basics}).  We then discuss $N =4$ basis functions, where one needs to have intermediate angular momenta to fully fix the coupling structure (\S\ref{sec:interm}). We next move to $N = 5$ functions, which have the same challenge (\S\ref{sec:N5}). We then (\S\ref{sec:odd}) outline how to obtain generating functions for the odd-parity isotropic basis functions, which is a more complicated matter and requires additional work beyond what is needed for the even-parity case. We then turn to expanding the Dirac Delta function in this basis (\S\ref{sec:dirac}), and subsequently present connections of the generating function to spherical Bessel function overlap integrals (\S\ref{sec:overlap}). In \S\ref{sec:alg} we suggest one possible application of the Cartesian forms of the isotropic functions, outlining a new algorithm to compute the N-Point Correlation Function (NPCF) of a given density field; this approach may offer a speed-up over the spherical-harmonic-based \textsc{encore} code \citep{encore} when run on a CPU. \S\ref{sec:concs} concludes.

\section{Basic Expansion and Generalized Gaunt Integrals}
\label{sec:basics}
\subsection{Definitions of the Isotropic Basis Functions}
For completeness, we first define  our target, the isotropic basis functions; the definition is taken from \cite{Cahn}, where they were first developed. In \cite{Cahn} \S2 the definition  is  written in a  compact and general way; here  we write all definitions out explicitly for clarity. In this work, we pursue up to $N = 5$ argument  basis functions. For $N=2$, we have
\begin{align}
    \mathcal{P}_{\ell \ell}(\hat{r}_1, \hat{r}_2) \equiv \frac{(-1)^{\ell}}{\sqrt{2\ell + 1}} \sum_m Y_{\ell m}(\hat{r}_1)
    Y_{\ell m}^*(\hat{r}_2).
    \label{eqn:iso_2}
\end{align}
For $N = 3$ we have
\begin{align}
\mathcal{P}_{\ell_1 \ell_2 \ell_3} 
(\hat{r}_1, \hat{r}_2, \hat{r}_3) 
\equiv (-1)^{\ell_1  + \ell_2  +  \ell_3} \sum_{m_1, m_2,  m_3} 
\begin{pmatrix}
\ell_1 & \ell_2 & \ell_3\\
m_1 & m_2 & m_3
\end{pmatrix}
Y_{\ell_1 m_1}(\hat{r}_1)
    Y_{\ell_2 m_2}(\hat{r}_2)
    Y_{\ell_3 m_3}(\hat{r}_3).
    \label{eqn:iso_3}
\end{align}
For $N=4$, we have
\begin{align}
    \mathcal{P}_{\ell_1  \ell_2 (\ell_{12})\ell_3 \ell_4}
    (\hat{r}_1, \hat{r}_2, \hat{r}_3,\hat{r}_4)  &=  (-1)^{\ell_1 + \ell_2+ \ell_3+ \ell_4} (-1)^{\ell_{12}}\sqrt{2\ell_{12}  + 1}
    \sum_{m_1, \ldots, m_4, m_{12}} 
    \begin{pmatrix}
\ell_1 & \ell_2 & \ell_{12}\\
m_1 & m_2 & -m_{12}
\end{pmatrix}\nonumber\\
&\times \begin{pmatrix}
\ell_{12} & \ell_3 & \ell_4\\
m_{12} & m_3 & m_4
\end{pmatrix}Y_{\ell_1 m_1}(\hat{r}_1)
    Y_{\ell_2 m_2}(\hat{r}_2)
    Y_{\ell_3 m_3}(\hat{r}_3)
    Y_{\ell_4 m_4}(\hat{r}_4).
    \label{eqn:iso_4}
\end{align}
For $N=5$, we have
\begin{align}
    \label{eqn:iso_5}
    &\mathcal{P}_{\ell_1  \ell_2(\ell_{12}) \ell_3 (\ell_{123}) \ell_4\ell_5}
    (\hat{r}_1, \hat{r}_2, \hat{r}_3,\hat{r}_4, \hat{r}_5)  =  (-1)^{\ell_1 + \ell_2+ \ell_3+ \ell_4 +  \ell_5} (-1)^{\ell_{12} + \ell_{123}}\sqrt{2\ell_{12}  + 1}
    \sqrt{2\ell_{123}  + 1}\nonumber\\
    &\times \sum_{m_1, \ldots, m_5, m_{12}, m_{123}} 
    \begin{pmatrix}
\ell_1 & \ell_2 & \ell_{12}\\
m_1 & m_2 & -m_{12}
\end{pmatrix} \begin{pmatrix}
\ell_{12} & \ell_3 & \ell_{123}\\
m_{12} & m_3 & -m_{123}
\end{pmatrix}
\begin{pmatrix}
\ell_{123} & \ell_4 & \ell_{5}\\
m_{123} & m_4 & m_5
\end{pmatrix}\\
&\times Y_{\ell_1 m_1}(\hat{r}_1)
    Y_{\ell_2 m_2}(\hat{r}_2)
    Y_{\ell_3 m_3}(\hat{r}_3)
    Y_{\ell_4 m_4}(\hat{r}_4)
    Y_{\ell_5 m_5}(\hat{r}_5).\nonumber
\end{align}

\subsection{Using the Plane Wave Expansion}
Consider a vector 
\begin{align}
\vec{R} \equiv \sum_{i = 1}^N \vec{r}_i,
\label{eqn:R_Def}
\end{align} 
and a momentum $\vec{p}$ dual to $\vec{R}$. We note that, for what follows, it is not essential that $\vec{p}$ be a momentum; it can be simply considered as a dummy parameter if one prefers. Let us now examine two different ways of writing the plane wave expansion into spherical harmonics and spherical Bessel functions. First, we have
\begin{align}
e^{i \vec{p} \cdot \vec{R}} = 4\pi \sum_{\ell m} i^{\ell} j_{\ell}(pR) Y_{\ell m}^*(\hat{p}) Y_{\ell m} (\hat{R}).
\label{eqn:pwe}
\end{align}
We note that here and in all that follows, if the limits of an orbital angular-momentum-like index (\textit{e.g.} $\ell$ above) are not indicated explicitly, they are from zero to infinity. Throughout, we will refer to $m$-like indices as projective quantum numbers; they are simply the $z$-component of the orbital angular momentum. If  the limits  of an $m$-like index are not indicated they are from $-\ell$ to $+\ell$, inclusive. 

Now let us write equation (\ref{eqn:pwe}) out by factoring the plane wave as $\exp[i \vec{p} \cdot \vec{r}_1] \times \cdots \times \exp[i \vec{p} \cdot \vec{r}_N]$. We then have
\begin{align}
e^{i \vec{p} \cdot \vec{R}} =  (4\pi)^N \prod_{{i=1}}^{{N}}  \left\{\sum_{\ell_i}  i^{\ell_i} j_{\ell_i}(pr_i)  \sum_{m_i} Y_{\ell_i m_i}^*(\hat{p})Y_{\ell_i m_i} (\hat{r}_i) \right\}.
\end{align}
 
We now integrate each of the above equations against $Y_{\ell m}(\hat{p})$, and set them equal. We have
\begin{align} 
\boxed{
4\pi i^{\ell} j_{\ell}(pR) Y_{\ell m} (\hat{R}) = (4\pi)^N \sum_{\ell_{i=1,\ldots,N}} i^{\sum_{{i=1}}^{{N}} \ell_i}\prod_{{i=1}}^{{N}}\left\{ j_{\ell_i}(pr_i)  \sum_{m_i} Y_{\ell_i m_i} (\hat{r}_i) \right\} \int d\Omega_p\; Y_{\ell m}(\hat{p}) \prod_{i=1}^N  Y_{\ell_i m_i}^*(\hat{p}).}
\label{eqn:main_3} 
\end{align}
We note that each $\ell_i$ assumes a full range of values, over which we are summing. We now convert $Y^{{*}}_{\ell m}(\hat{p}) = (-1)^m Y_{\ell -m}(\hat{p})$ so that our $d\Omega_p$ integral is over all conjugated spherical harmonics, in which case it is then equal to what it would be had they been all unconjugated (this equality follows from the fact that the result must be real).\footnote{$d\Omega_p$ is the integration measure for integrating over all directions of $\hat{p}$.} We then may rewrite this integral as
\begin{align}
\label{eqn:gen_gaunt}
\int d\Omega_p\; Y_{\ell m}(\hat{p}) \prod_{i = 1}^N  Y_{\ell_i m_i}^*(\hat{p}) \equiv (-1)^m \mathcal{G}_{\ell_1 \cdots \ell_N \ell}^{m_1 \cdots m_N -m},
\end{align}
where $\mathcal{G}$ is the generalized Gaunt integral over $(N+1)$ all unconjugated spherical harmonics with the indicated angular momenta and projective quantum numbers. The last projective quantum number in the superscript is $-m$, not a subtraction of $m$ from $m_N$. The generalized Gaunt can be considered to be defined by this relation.


\subsection{Generalized Gaunt Integrals}
We will explicitly write down this integral for $N= 2, 3, 4$ and $5$. We note the $N$ here is the number of arguments of the basis functions we are studying, \textit{i.e.} the number of $\hat{r}_i$; the number of harmonics involved in the integral (\ref{eqn:gen_gaunt}) is then $N + 1$, as one may see because there is one additional harmonic beyond the product indexed by $i$.  

\subsubsection{$N = 2$ Argument Basis Functions $\rightarrow$ 3 Harmonics}
For $N=2$, we thus have three spherical harmonics, so: 
\begin{align}
\mathcal{G}_{\ell_1 \ell_2 \ell}^{m_1 m_2 -m} = \mathcal{C}_{\ell_1 \ell_2 \ell} \begin{pmatrix}
\ell_1 & \ell_2 & \ell\\
0 & 0 & 0
\end{pmatrix}
\begin{pmatrix}
\ell_1 & \ell_2 & \ell\\
m_1 & m_2 & -m
\end{pmatrix},
\label{eqn:near_calC}
\end{align}
with 
\begin{align}
\label{eqn:calC}
\mathcal{C}_{\ell_1 \ell_2 \ell} \equiv \sqrt{\frac{(2\ell_1 + 1) (2\ell_2 + 1) (2\ell + 1)}{4\pi}}. 
\end{align}

\subsubsection{$N = 3$ Argument Basis Functions $\rightarrow$ 4 Harmonics}
For $N = 3$ we have, choosing the same coupling approach (``pyramidal'') as in \cite{Cahn}, and projecting the product of $Y_{\ell_1 m_1} (\hat{p}) Y_{\ell_2 m_2}  (\hat{p})$ onto a sum over $Y_{\ell_{12} m_{12}}(\hat{p}) $ (where this last is unconjugated) using \cite{nist} 34.3.20, we find
\begin{align}
\mathcal{G}_{\ell_1 \ell_2 \ell_3 \ell}^{m_1 m_2 m_3 -m} = \sum_{\ell_{12}  } (-1)^{m_{12}} \mathcal{G}_{\ell_1 \ell_2 \ell_{12}}^{m_1 m_2 -m_{12}} \mathcal{G}_{\ell_{12} \ell_3 \ell}^{m_{12} m_3 -m}.
\end{align}
We do not need to sum over $m_{12}$ on the right-hand side because $m_1 + m_2 - m_{12} = 0$ fixes $m_{12}$. This constraint stems from the fact that the usual Gaunt integrals contain a 3-$j$ symbol whose bottom row is these projective quantum numbers.

\subsubsection{$N = 4$ Argument Basis Functions $\rightarrow$ 5 Harmonics}
For $N= 4$, we have
\begin{align}
\mathcal{G}_{\ell_1 \ell_2 \ell_3 \ell_4 \ell}^{m_1 m_2 m_3 m_4 -m} = \sum_{\ell_{12}, \ell_{123}} (-1)^{m_{12}} \mathcal{G}_{\ell_1 \ell_2 \ell_{12}}^{m_1 m_2 -m_{12}} (-1)^{m_{123}} \mathcal{G}_{\ell_{12} \ell_3 \ell_{123}}^{m_{12} m_3 -m_{123}} \mathcal{G}_{\ell_{123} \ell_4 \ell}^{m_{123} m_4 -m}.
\end{align}
Again we note there is no need for a sum over any multi-index projective quantum numbers (\textit{e.g.} $m_{12}$ or similar).

\subsubsection{$N = 5$ Argument Basis Functions $\rightarrow$ 6 Harmonics}
For $N = 5$ we have
\begin{align}
\mathcal{G}_{\ell_1 \ell_2 \ell_3 \ell_4 \ell_5 \ell}^{m_1 m_2 m_3 m_4 m_5 -m} = &\sum_{\ell_{12}, \ell_{123}, \ell_{1234}} \qquad  (-1)^{m_{12}} \mathcal{G}_{\ell_1 \ell_2 \ell_{12}}^{m_1 m_2 -m_{12}} (-1)^{m_{123}} \mathcal{G}_{\ell_{12} \ell_3 \ell_{123}}^{m_{12} m_3 -m_{123}} (-1)^{m_{1234}}\nonumber\\
&\qquad \times \mathcal{G}_{\ell_{123} \ell_4 \ell_{1234}}^{m_{123} m_4 -m_{1234}} \mathcal{G}_{\ell_{1234} \ell_5 \ell}^{m_{1234} m_5 -m}.
\end{align}

\subsubsection{Use in Fundamental Relation}
With these generalized Gaunt integrals in hand, we now return to our main thread, the fundamental, boxed relation (\ref{eqn:main_3}). With the integration over $d\Omega_p$ now performed (we recall equation \ref{eqn:gen_gaunt}), it becomes:
\begin{align}
i^{\ell} j_{\ell}(pR) Y_{\ell m} (\hat{R}) = (4\pi)^{N-1} \sum_{\ell_{i=1,\ldots, N}} i^{\sum_{i=1}^{N}} {\ell_i}\prod_{{i=1}}^{{N}}\left\{ j_{\ell_i}(pr_i)  \sum_{m_i} Y_{\ell_i m_i} (\hat{r}_i) \right\} (-1)^m \mathcal{G}_{\ell_1 \cdots \ell_N \ell}^{m_1 \cdots m_N -m}.
\end{align}

\subsubsection{Reduction when $\ell = 0$} 
Let us now consider the case where $\ell \to 0$, meaning $m=0$ as well. Doing so is motivated by the fact that we are seeking isotropic basis functions of the collection of unit vectors $\hat{r}_i$; hence, the way we have set our relations up, they must have coupled to an isotropic mode in $\hat{p}$ prior to integration over it. We may impose this condition by taking it that $\ell = 0$ in the harmonic in $\hat{p}$ in equation (\ref{eqn:main_3}).\footnote{We might have reached this point more directly by simply integrating over $d\Omega_p$ in equation (\ref{eqn:main_3}), which is of course equivalent to integrating against $1 = \sqrt{4\pi}\; Y_{00}^*(\hat{p})$. The approach used above, however, is more general. In particular, it can actually offer a set of ``anisotropic'' basis functions as well if $\ell$ is allowed to be non-zero (\textit{e.g.} as noted in \cite{philcox_hi_d}; see also \cite{aniso_3}).} We obtain
\begin{align}
\boxed{
\frac{1}{\sqrt{4\pi}} j_0 (pR)  = (4\pi)^{N-1} \sum_{{\ell_{i=1,\ldots, N}}} i^{\sum_{{i=1}}^{{N}} \ell_i}\prod_{{i=1}}^{{N}}\left\{ j_{\ell_i}(pr_i)  \sum_{m_i} Y_{\ell_i m_i} (\hat{r}_i) \right\} \mathcal{G}_{\ell_1 \cdots \ell_N 0}^{m_1 \cdots m_N 0}.}
\label{eqn:fund}
\end{align}

\subsection{Functions of Two Arguments: $N=2$}
The generalized Gaunt integral becomes, for $N=2$ (recalling equation \ref{eqn:near_calC}):
\begin{align} 
\mathcal{G}_{\ell_1 \ell_2 0}^{m_1 m_2 0} = \frac{1}{\sqrt{4\pi}}\;(-1)^{m_1} \delta^{\rm K}_{\ell_1 \ell_2} \delta^{\rm K}_{m_1 -m_2},
\end{align}
obtained by applying NIST DLMF 34.3.1 twice. In particular, we used that a 3-$j$ symbol with zeros in the third column forces equality between the first and second angular momenta, and that the associated two projective quantum numbers sum to zero. 

Hence our $N=2$ result becomes
\begin{align}
 j_0 (pR)  = 4\pi \sum_{\ell_1 } (-1)^{\ell_1} j_{\ell_1} (pr_1) j_{\ell_1}(pr_2)  \sum_{m_1} Y_{\ell_1 m_1} (\hat{r}_1) Y^*_{\ell_1 m_1} (\hat{r}_2).
 \end{align}
 We used the conjugate identity to absorb the projective quantum number-dependent phase into the second harmonic above, and we note that we set $m = 0$ as implied by $\ell = 0$. We note that the left-hand side depends on the angles between $\hat{r}_1$ and $\hat{r}_2$ because $R=|\bf{r}_1+\bf{r}_2|$.
 
 Comparing with equation (\ref{eqn:iso_2}), we have a fundamental relation for generating the $N = 2$ isotropic basis functions, as 
\begin{align}
\boxed{
 j_0 (pR)  = 4\pi  \sum_{\ell_1 } \sqrt{2\ell_1 +1} \;  j_{\ell_1} (pr_1) j_{\ell_1}(pr_2)  \mathcal{P}_{\ell_1 \ell_1}(\hat{r}_1, \hat{r}_2).}
 \label{eqn:2_fun}
 \end{align} 
We may now take the limit that $p \to 0$, and using that $j_0(x) = \sin x/x$, we have the series for the left-hand side as 
\begin{align}
\label{eqn:j0_exp}
\boxed{
    j_0(pR) \approx 1 - \frac{(pR)^2}{6} + \frac{(pR)^4}{120} - \frac{(pR)^6}{5,040} + \cdots.}
\end{align}
We box this relation because it will be of much use in what follows. On the right-hand side, we have the series for each spherical Bessel function (sBf) at lowest order as $(pr_i)^{\ell_1}/(2\ell_1 + 1)!!$. The double factorial $n!!$ is the product of all the integers from 1 up to $n$ (inclusive) that have the same parity as $n$.\footnote{\textit{E.g.} \url{https://mathworld.wolfram.com/DoubleFactorial.html}.}

We may now demand term-by-term equality for the coefficients of powers of $p$ on each side of equation (\ref{eqn:2_fun}); hence the above relation shows that $j_0(pR)$ with $R \equiv |\vec{r}_1 + \vec{r}_2|$ is a generating function for the $N=2$ isotropic basis functions.\footnote{There is no approximation involved in this, even though we used a series for the sBfs. It is a power-matching argument, just like that typically employed in solving differential equations via the method of Frobenius. Fundamentally, it is because, roughly, the powers of $pR$ may be considered an orthogonal basis, which is, roughly, the result of the Stone-Weierstrass theorem. }
\subsection{Examples for $N=2$}
We may easily read off results for the first few $\mathcal{P}_{\ell_1  \ell_1}$. Throughout this section, we will make use of equation (\ref{eqn:j0_exp}) for the left-hand side of equation (\ref{eqn:2_fun}).  

For $\ell_1 = 0$, we have after a line of algebra that $\mathcal{P}_{00} = 1/(4\pi)$. For $\ell_1 = 1$, we expand $|\vec{r}_1 + \vec{r}_2|^2$ on the left-hand side and retain only the part proportional to $r_1 r_2$, to match the right-hand side. We find $\mathcal{P}_{11}(\hat{r}_1, \hat{r}_2) = -\sqrt{3} /(4\pi)\; (\hat{r}_1 \cdot \hat{r}_2)$. 

For $\ell_1 = 2$, there is a subtlety. The sub-leading expansion of the $\ell_1 = 0$ term on the right-hand side can contribute a term in $p^4$ (the relevant power for $\ell_1 = 2$) as well. This contribution also will enter proportional to $r_1^2 r_2^2$, so we cannot eliminate it by selecting based on the $r_i$-dependence.\footnote{The $\ell_1 = 1$ terms can contribute as $p^4$ but, given the parity of each $j_1$, never with dependence on $r_1^2 r_2^2$; an even-parity term in $p$ will have to have mismatched powers of $r_1$ and $r_2$.} We also note that no $\ell_1>2$ can contribute such a low power, given the leading-order behavior of the sBfs at small argument. 

However, we must indeed count the contribution of the sub-leading $\ell_1 = 0$; it must be subtracted off to isolate the contribution to the right-hand side sum due only to $\mathcal{P}_{22}$. We have, focusing only on terms proportional to $p^4$ and $r_1^2 r_2^2$, 
\begin{align}
    \frac{R^4}{120}\bigg|_{r_1^2 r_2^2} = 4\pi \left[\frac{1}{36} \mathcal{P}_{00}(\hat{r}_1, \hat{r}_2)  + \frac{\sqrt{5}}{15^2}\mathcal{P}_{22}(\hat{r}_1, \hat{r}_2)  \right]. 
\end{align}
Our notation on the left-hand side means that when $R^4$ is expanded only terms in $r_1^2 r_2^2$ should be retained (and we that we have divided the $r_1^2 r_2^2$ out from both sides). We find, working in terms of $\mu_{12} \equiv \hat{r}_1 \cdot \hat{r}_2$, that
\begin{align}
    \frac{2+4\mu_{12}^2}{120} = \frac{1}{36} + 4\pi \frac{\sqrt{5}}{15^2} \mathcal{P}_{2,2}(\hat{r}_1, \hat{r}_2);
\end{align}
solving for $\mathcal{P}_{2 2}$ yields
\begin{align}
    \mathcal{P}_{2 2}(\hat{r}_1, \hat{r}_2) = \frac{3\sqrt{5}}{8\pi}\left[ (\hat{r}_1 \cdot \hat{r}_2)^2 -\frac{1}{3}\right]
\end{align}
in agreement with \cite{Cahn} equation A.1., last line.

We observe that due to the power-matching in the $r_i$, we need only to consider one term from each series for lower $\ell$ than the $\ell_i$ under consideration, and we also need only to consider expansions of lower $\ell$ sBfs of the same parity here.

Working through $\ell_1 = 3$, we find the fundamental equation
\begin{align}
    -\frac{R^6}{5,040}\bigg|_{r_1^3 r_2^3} = 4\pi \left[ \frac{\sqrt{3}}{30^2} \mathcal{P}_{1,1}(\hat{r}_1, \hat{r}_2) + \frac{\sqrt{7}}{105^2}\mathcal{P}_{3,3}(\hat{r}_1, \hat{r}_2)\right].
\end{align}
Expanding the left-hand side gives
\begin{align}
    -\frac{R^6}{5,040}\bigg|_{r_1^3 r_2^3} = -\frac{1}{5,040} \left[8\mu_{12}^3 + 12\mu_{12}\right].
\end{align}
Solving for $\mathcal{P}_{3 3}$ we obtain
\begin{align}
    \mathcal{P}_{3 3}(\hat{r}_1, \hat{r}_2) = -\frac{\sqrt{7}}{4\pi} \mathcal{L}_3(\hat{r}_1 \cdot \hat{r}_2),
\end{align}
where $\mathcal{L}_3$ is the Legendre polynomial of order three. This result is in agreement with \cite{Cahn} equation 3 with $\Lambda = 3$ there.

\subsection{Functions of Three Arguments: $N=3$}
We now seek to extend our work to $N>2$ functions. For $N=3$, the generalized Gaunt integral becomes
\begin{align}
\mathcal{G}_{\ell_1 \ell_2 \ell_3 0}^{m_1 m_2 m_3 0} = \frac{1}{\sqrt{4\pi}} \mathcal{G}_{\ell_1 \ell_2 \ell_3}^{m_1 m_2 m_3}.
\end{align}
Generally, we note that setting the last angular momentum and projective quantum number to $``00"$ always introduces a factor of $ 1/\sqrt{4\pi}$.

Using this in our fundamental relation (\ref{eqn:fund}), we find after some simplification
\begin{align}
j_0(pR) &= (4 \pi)^2 \sum_{\ell_1 \ell_2 \ell_3} i^{\ell_1 +\ell_2 +\ell_3} j_{\ell_1}(pr_1) j_{\ell_2}(pr_2) j_{\ell_3}(pr_3)  \begin{pmatrix}
\ell_1 & \ell_2 & \ell_3\\
0 & 0 & 0
\end{pmatrix} \mathcal{C}_{\ell_1 \ell_2 \ell_3} \nonumber\\
&\qquad\times \sum_{m_1 m_2 m_3}  \begin{pmatrix}
\ell_1 & \ell_2 & \ell_3\\
m_1 & m_2 & m_3
\end{pmatrix}
Y_{\ell_1 m_1}(\hat{r}_1) Y_{\ell_2 m_2}(\hat{r}_2) Y_{\ell_3 m_3}(\hat{r}_3).
\end{align}
Using equation (\ref{eqn:iso_3}) this becomes
\begin{align}
\boxed{
j_0(pR) = (4 \pi)^2 \sum_{\ell_1 \ell_2 \ell_3} i^{\ell_1 +\ell_2 +\ell_3} 
j_{\ell_1}(pr_1) j_{\ell_2}(pr_2) j_{\ell_3}(pr_3)  
\begin{pmatrix}
\ell_1 & \ell_2 & \ell_3\\
0 & 0 & 0
\end{pmatrix} 
\mathcal{C}_{\ell_1 \ell_2 \ell_3}  \mathcal{P}_{\ell_1 \ell_2 \ell_3}(\hat{r}_1, \hat{r}_2, \hat{r}_3).}
\label{eqn:sixteen}
\end{align}
We note that the right-hand side will always be real, since the zero-projective quantum number 3-$j$ symbol guarantees that $\ell_1 + \ell_2 + \ell_3$ is even. We recall that $\mathcal{C}_{\ell_1 \ell_2 \ell_3}$ is defined by equation (\ref{eqn:calC}). 

Now, let us consider the structure above further. A given power of $p$ on the left-hand side (when we expand in a series in $p$) must be equal to $\ell_1 + \ell_2 + \ell_3$. One can think of this as analogous to an overall energy level (as in the usual model of an atom, which has overall energy levels and then shells within them, \textit{s p d f} \textit{etc.}) for the set of isotropic functions satisfying the constraint. Then, within each level, we may select out one or another isotropic function by requiring a given behavior in $r_1, r_2, r_3$. 

This latter aspect is analogous to how the Legendre polynomials are organized when written using their generating function as in equation (\ref{eqn:multip}). There, one requires a given behavior in $r_</r_>$ at each order. For Legendres, since they have only two arguments, they have only \textit{one} $\ell$ (as we indeed saw in our $N=2$ isotropic functions above), so there is no distinction, and hence no need for ordering, within a given ``energy level''---there is only one ``shell'' in each. Here, when we go to $N=3$, we have multiple ``shells'' within each ``energy level'', and we can use $p$ to indicate the energy level and then the power of each of the $r_i$ to indicate the shell.

We make one final remark. $j_0(pR)$ is even-parity, and hence only even powers of $p$ will enter the expansion for $p\ll 1$. Consequently, we must have that $\ell_1 + \ell_2 + \ell_3$ is even and so we can only have even-parity isotropic functions on the right-hand side. This is also enforced by the even-sum rule on the zero-projective quantum number 3-$j$ symbol that appears. We turn to the issue of odd-parity basis functions in \S\ref{sec:odd}.

\subsection{Examples for $N=3$}
Here we work out several examples of the three-argument basis functions using this work's formalism. We begin with $\mathcal{P}_{110}$. Here, we have
\begin{align}
    -\frac{(pR)^2}{6}\bigg|_{r_1 r_2} = (4\pi)^2 \left[ - \frac{1}{9} \left(-\frac{1}{\sqrt{3}}\right) \frac{3}{\sqrt{4\pi}} \right]\mathcal{P}_{110}(\hat{r}_1, \hat{r}_2, \hat{r}_3),
\end{align}
where we used equation (\ref{eqn:j0_exp}) for the left-hand side and equation (\ref{eqn:sixteen}) for the right-hand side, evaluating the phase and 3-$j$ symbol with $\ell_1 = \ell_2 = 1$ and $\ell_3 = 0$, and using the definition of $\mathcal{C}_{\ell_1 \ell_2 \ell_3}$ equation (\ref{eqn:calC}). We recall that our notation $\bigg|_{r_1 r_2}$ means that we have taken only terms involving this set of powers of the $r_i$ and that we have then cancelled any powers of $p$ and the $r_i$ from both sides.

Now expanding $R^2 = |\vec{r}_1 + \vec{r}_2|^2$ and retaining only the terms in $r_1 r_2$, a few lines of algebra shows that 
\begin{align}
\label{eqn:110}
    \mathcal{P}_{110}(\hat{r}_1, \hat{r}_2, \hat{r}_3) = -\frac{1}{4\pi} \sqrt{\frac{3}{4\pi}}\; \left(\hat{r}_1 \cdot \hat{r}_2 \right),
\end{align}
in agreement with \cite{Cahn} Appendix A.2., first line.

We now pursue a case requiring subtraction of sub-leading terms in the series of lower-order sBfs on the right-hand side, as we already saw was required in the $N = 2$ functions. We consider $\mathcal{P}_{112}$. We have, using equation (\ref{eqn:j0_exp}) for the left-hand side and equation (\ref{eqn:sixteen}) for the right-hand side, that
\begin{align}
\label{eqn:112}
    \frac{(pR)^4}{120}\bigg|_{r_1 r_2 r_3^2} = (4\pi)^2 \left[-\frac{1}{54} \sqrt{\frac{3}{4\pi}}\; \mathcal{P}_{110}+ \frac{1}{45} \sqrt{\frac{2}{3(4\pi)}}\; \mathcal{P}_{{112}} \right],
\end{align}
where the first term on the right-hand side came from noticing that expanding $j_1(pr_1) j_1(pr_2)$ to leading order each, but $j_0(p r_3)$ to second order, will also give an overall $p^4$. The second term is simply from taking the lowest-order terms in the series for, respectively, $j_1(pr_1), j_1(pr_2)$ and $j_2(pr_3)$.

Computing the expansion of $R^4$ and retaining only the term indicated by the evaluation bar on the left-hand side, we find
\begin{align}
    \frac{(pR)^4}{120}\bigg|_{r_1 r_2 r_3^2} = \frac{1}{30}\left[ \hat{r}_1 \cdot \hat{r}_2 + 2 (\hat{r}_2 \cdot \hat{r}_3) (\hat{r}_3 \cdot \hat{r}_1)\right].
\end{align}
Inserting this result in equation (\ref{eqn:112}), using our result for $\mathcal{P}_{110}$ from equation (\ref{eqn:110}, and solving for $\mathcal{P}_{112}$ yields
\begin{align}
    \mathcal{P}_{112}(\hat{r}_1, \hat{r}_2, \hat{r}_3) = \sqrt{\frac{27}{2(4\pi)^3}}\left[(\hat{r}_1 \cdot \hat{r}_3)(\hat{r}_2 \cdot \hat{r}_3) - \frac{1}{3} (\hat{r}_1 \cdot \hat{r}_2) \right],
\end{align}
in agreement with \cite{Cahn} Appendix A.2., third line. 

\section{Dealing with Intermediate Angular Momenta: $N = 4$}
\label{sec:interm}
We here discuss how to deal with the intermediate angular momenta required as one goes to isotropic functions of four arguments or more. We first show that the most obvious extension of the approach of the previous sections fails, and then outline the correct method.
\subsection{Failure of the Naive Approach}
We now consider the generalized Gaunt for $N=4$. Again making use of NIST DLMF 34.3.1 and after some simplification, we have
\begin{align}
\mathcal{G}_{\ell_1 \ell_2 \ell_3 \ell_4 0}^{m_1 m_2 m_3 m_4 0} = \frac{1}{\sqrt{4\pi}} \sum_{\ell_{12}} (-1)^{m_{12}} \mathcal{G}_{\ell_1 \ell_2 \ell_{12}}^{m_1 m_2 -m_{12}} \mathcal{G}_{\ell_{12} \ell_3 \ell_4}^{m_{12} m_3 m_4}.
\end{align}
Inserting this in our fundamental relation equation (\ref{eqn:fund}), we find
\begin{align}
j_0 (pR)  = (4\pi)^3 \sum_{\ell_1,\ldots, \ell_4} i^{\ell_1 + \ell_2 + \ell_3 + \ell_4}\prod_{i=1}^4 \left\{ j_{\ell_i}(pr_i)  \sum_{m_i} Y_{\ell_i m_i} (\hat{r}_i) \right\}  \mathcal{G}_{\ell_1 \ell_2 \ell_3 \ell_4 0}^{m_1 m_2 m_3 m_4 0}.
\end{align}
Using \cite{Cahn} equation (12), we have
\begin{align}
j_0 (pR)  = &(4\pi)^2 \sum_{\ell_1,\ldots, \ell_4} i^{\ell_1 + \ell_2 + \ell_3 + \ell_4} \left[ \prod_{i=1}^4  j_{\ell_i}(pr_i) \right] (-1)^{\ell_1 + \ell_2 +\ell_3 + \ell_4} \mathcal{D}^{\rm P}\nonumber\\
&\qquad\qquad\qquad \times \sum_{\ell_{12}} \sqrt{2\ell_{12} + 1}
\begin{pmatrix}
\ell_1 & \ell_2 & \ell_{12}\\
0 & 0 & 0
\end{pmatrix}
\begin{pmatrix}
\ell_{12} & \ell_3 & \ell_4\\
0 & 0 & 0
\end{pmatrix}
(-1)^{\ell_{12}} \nonumber\\
&\qquad\qquad\qquad \times \mathcal{P}_{\ell_1 \ell_2 (\ell_{12}) \ell_3 \ell_4} (\hat{r}_1, \hat{r}_2, \hat{r}_3, \hat{r}_4).
\label{eqn:naive_4}
\end{align}
The additional phase $(-1)^{\ell_1 + \ell_2 + \ell_3 + \ell_4}$ comes from the definition of $\mathcal{P}_{\Lambda}$, \cite{Cahn} equation 11, in particular $\epsilon(\Lambda)$ as defined in equation 9 there;
$\Lambda \equiv \{ \ell_1, \ell_2, \ell_{12}, \ldots  \}$, \textit{i.e.} it is a specification of both the primary and intermediate angular momenta. The symbol $\mathcal{D}^{\rm P}$ is a product of $\sqrt{2\ell_i+1}$ over all \textit{primary} angular momenta (defined in \cite{Cahn} equation 35). Briefly, a primary angular momentum is one that corresponds to a spherical harmonic entering the expansion; we always denote them with a single numerical subscript, \textit{e.g.} $\ell_1$. An \textit{intermediate} angular momentum is one representing the coupling of either two primaries, or a primary and an intermediate, and always will appear with a subscript involving multiple numbers, \textit{e.g.} $\ell_{12}, \ell_{123}$, \textit{etc}. The intermediates do not have spherical harmonics associated with them. The need for intermediates is discussed more fully in \cite{Cahn}; see also \cite{const4} and \cite{hu_01}.

We now observe that expanding both sides in $p$ and requiring power-by-power equality as we have in the previous cases, we will have no way to pick out any particular $\ell_{12}$, but will be left with the sum over all of those allowed. This is because the sBfs on the right-hand side have no index $\ell_{12}$. 

\subsection{Picking Out the Intermediates}
The failure of the naive approach above means we require another way to proceed. 

Since each of our angular momenta before were associated with a unit vector $\hat{p}$ that is integrated out, we now consider also associating the intermediate angular momentum with a vector, which we denote $\vec{u}_{12}$. Let us examine the expansion
\begin{align}
&e^{i \vec{p} \cdot  (\vec{r}_1 + \vec{r}_2)} e^{i \vec{p} \cdot \vec{u}_{12}} e^{i \vec{p}' \cdot (\vec{r}_3 + \vec{r}_4)} e^{i \vec{p}' \cdot \vec{u}_{12}}
= (4\pi)^4 \sum_{\ell \ell'} \;\sum_{\ell_{12} \ell_{12}'} \; \sum_{{\rm all}\; m} i^{\ell + \ell' + \ell_{12} + \ell_{12}'} j_{\ell}(pR) j_{\ell'}(p'R') Y_{\ell m} (\hat{p}) Y_{\ell m}^* (\hat{R})   \nonumber\\
&\times Y_{\ell' m'} (\hat{p}')Y_{\ell' m'}^* (\hat{R}') j_{\ell_{12}}(p u_{12}) j_{\ell'_{12}}(p' u_{12}) Y_{\ell_{12} m_{12}} (\hat{p}) Y_{\ell_{12} m_{12}}^*(\hat{u}_{12}) Y_{\ell_{12}' m_{12}'} (\hat{p}') Y_{\ell_{12}' m_{12}'}^*(\hat{u}_{12}),
\label{eqn:N4}
\end{align}
where $\vec{R} \equiv \vec{r}_1 + \vec{r}_2$ and $\vec{R'} \equiv \vec{r}_3 + \vec{r}_4$. Integrating equation (\ref{eqn:N4}) over $d\Omega_p d\Omega_{p'}$ gives Kronecker deltas on the right-hand side as 
\begin{align}
\label{eqn:kron-4}
\int d\Omega_p d\Omega_{p'} 
Y_{\ell m} (\hat{p})  Y_{\ell_{12} m_{12}} (\hat{p}) 
Y_{\ell' m'} (\hat{p}')
Y_{\ell_{12}' m_{12}'} (\hat {p}') =
\delta^{\rm K}_{\ell \ell_{12}} \delta^{\rm K}_{m -m_{12}} \delta^{\rm K}_{\ell' \ell'_{12}} \delta^{\rm K}_{m' -m_{12}'} (-1)^{m_{12} + m'_{12}},
\end{align}
which will eliminate several sums. We used that\footnote{\url{https://dlmf.nist.gov/14.30}} $Y_{\ell m} = (-1)^m Y_{\ell -m}^*$ and then the spherical harmonics' orthogonality relation.  

We then integrate our result again, over $d\Omega_{12}$. After simplifying what results and then applying the spherical harmonic addition theorem to convert it from a product of harmonics into a Legendre polynomial, we finally find
\begin{align}
&\int d\Omega_p d\Omega_{p'} d\Omega_{12} \; e^{i \vec{p} \cdot  (\vec{r}_1 + \vec{r}_2)} e^{i \vec{p} \cdot \vec{u}_{12}} e^{i \vec{p}' \cdot (\vec{r}_3 + \vec{r}_4)} e^{i \vec{p}' \cdot \vec{u}_{12}}\nonumber\\
&= (4\pi)^3 \sum_{\ell} (2 \ell + 1) \mathcal{L}_{\ell} (\hat{R} \cdot \hat{R}') j_{\ell}(pR) j_{\ell}(p'R') j_{\ell}(p u_{12}) j_{\ell}(p' u_{12}).
\label{eqn:lhs_four}
\end{align}
We note that the phase from equation (\ref{eqn:N4}) cancelled because $\ell + \ell_{12} + \ell' + \ell_{12}' = 2(\ell_{12} + \ell'_{12})$ due to equation (\ref{eqn:kron-4}) and then the $d\Omega_{12}$ integration set $\ell_{12}' = \ell_{12}$, so the power of $i$ in equation (\ref{eqn:N4}) becomes an even number divisible by two.

We now seek to relate this set of plane waves to the $N=4$ isotropic basis functions. We have, by rewriting each plane wave of a sum as the product of plane waves prior to expanding, that
\begin{align}
\label{eqn:38}
&e^{i \vec{p} \cdot  (\vec{r}_1 + \vec{r}_2)} e^{i \vec{p} \cdot \vec{u}_{12}} e^{i \vec{p}' \cdot (\vec{r}_3 + \vec{r}_4)} e^{i \vec{p}' \cdot \vec{u}_{12}}\nonumber\\
&= (4\pi)^6 \sum_{\ell_1 \ell_2 \ell_{12}} \;\sum_{\ell_3 \ell_4 \ell_{12}'} \;\sum_{{\rm all\;}m} i^{\ell_1 + \ell_2 + \ell_{12} + \ell_3 + \ell_4 + \ell_{12}'} j_{\ell_1}(pr_1)
j_{\ell_2}(pr_2)
j_{\ell_{12}}(pu_{12})
j_{\ell_3}(p'r_3)
j_{\ell_4}(p'r_4)
j_{\ell_{12}'}(p'u_{12})\nonumber\\
&\qquad \times Y_{\ell_1 m_1}(\hat{p}) 
Y_{\ell_1 m_1}^*(\hat{r}_1) 
Y_{\ell_2 m_2}(\hat{p}) 
Y_{\ell_2 m_2}^*(\hat{r}_2) 
Y_{\ell_{12} m_{12}}(\hat{p}) 
Y_{\ell_{12} m_{12}}^*(\hat{u}_{12})
Y_{\ell_3 m_3}(\hat{p}') 
Y_{\ell_3 m_3}^*(\hat{r}_3) 
\nonumber\\
&\qquad \times Y_{\ell_4 m_4}(\hat{p}') 
Y_{\ell_4 m_4}^*(\hat{r}_4) 
Y_{\ell_{12}' m_{12}'}(\hat{p}') 
Y_{\ell_{12}' m_{12}'}^*(\hat{u}_{12}).
\end{align}
We now perform the same manipulations on this relation as we did on equation (\ref{eqn:N4}), to wit, integrating over $d\Omega_p d\Omega_{p'}d\Omega_{12}$. Integrating over $d\Omega_p d\Omega_{p'}$ replaces the product of spherical harmonics in $\ell_1, \ell_2, \ell_{12}$ with a single Gaunt integral, and likewise for those in $\ell_3, \ell_4, \ell_{12}'$. Integrating what results over $d\Omega_{12}$, gives a product of Kronecker deltas on the right-hand side as $\delta_{\ell_{12} \ell_{12}'}^{\rm K} \delta_{m_{12} -m_{12}'}^{\rm K} (-1)^{m_{12}}$.

Inserting these results and simplifying, we obtain
\begin{align}
&\int d\Omega_p d\Omega_{p'}d\Omega_{12}\; e^{i \vec{p} \cdot  (\vec{r}_1 + \vec{r}_2)} e^{i \vec{p} \cdot \vec{u}_{12}} e^{i \vec{p}' \cdot (\vec{r}_3 + \vec{r}_4)} e^{i \vec{p}' \cdot \vec{u}_{12}} = (4\pi)^6\nonumber\\
& \times \sum_{\ell_1 \ell_2 \ell_{12}} \;\sum_{\ell_3 \ell_4} \;\sum_{m_1 m_2 m_3}(-1)^{\ell_{12}} i^{\ell_1 + \ell_2 + \ell_3 + \ell_4} j_{\ell_1}(pr_1) j_{\ell_2}(pr_2) j_{\ell_3} (p'r_3) j_{\ell_4} (p' r_4) j_{\ell_{12}} (p u_{12}) j_{\ell_{12}} (p' u_{12}) \nonumber\\
&\qquad \times \mathcal{G}_{\ell_1 \ell_2 \ell_{12}}^{m_1 m_2 m_{12}} \mathcal{G}_{\ell_{12} \ell_3 \ell_4}^{-m_{12} m_3 m_4} (-1)^{m_{12}} 
Y_{\ell_1 m_1}^*(\hat{r}_1) 
Y_{\ell_2 m_2}^*(\hat{r}_2) 
Y_{\ell_3 m_3}^*(\hat{r}_3) 
Y_{\ell_4 m_4}^*(\hat{r}_4).
\label{eqn:40}
\end{align}
There is no sum over $m_{12}$ because the Gaunt integrals fix it; this is also the case for $m_4$. 

Inserting the definition of the $N = 4$ isotropic functions, equation (\ref{eqn:iso_4}), we find
\begin{align}
\label{eqn:41}
&\int d\Omega_p d\Omega_{p'}d\Omega_{12}\; e^{i \vec{p} \cdot  (\vec{r}_1 + \vec{r}_2)} e^{i \vec{p} \cdot \vec{u}_{12}} e^{i \vec{p}' \cdot (\vec{r}_3 + \vec{r}_4)} e^{i \vec{p}' \cdot \vec{u}_{12}}\\
&= (4\pi)^5 \sum_{\ell_1 \ell_2 \ell_{12}} \sum_{\ell_3 \ell_4}  (-1)^{(\ell_1 + \ell_2 + \ell_3 + \ell_4)/2}  j_{\ell_1}(pr_1) j_{\ell_2}(pr_2) j_{\ell_3} (p'r_3) j_{\ell_4} (p' r_4) j_{\ell_{12}} (p u_{12}) j_{\ell_{12}} (p' u_{12})
\mathcal{D}^{\rm P} \nonumber\\
&\qquad \times (-1)^{\ell_1 + \ell_2 + \ell_3 + \ell_4}\sqrt{2\ell_{12} + 1} \begin{pmatrix}
\ell_1 & \ell_2 & \ell_{12}\\
0 & 0 & 0
\end{pmatrix}
 \begin{pmatrix}
\ell_{12} & \ell_3 & \ell_4\\
0 & 0 & 0
\end{pmatrix}
\mathcal{P}_{\ell_1 \ell_2 (\ell_{12}) \ell_3 {\ell_4}}(\hat{r}_1, \hat{r}_2, \hat{r}_3, \hat{r}_4).\nonumber
\end{align}
We used that $\ell_1 + \ell_2 + \ell_{12}$ and $\ell_{12} + \ell_3 + \ell_4$ are even (due to the two zero-projective quantum number 3-$j$ symbols) to conclude that the parity of $\ell_{1} + \ell_2$ is that of $\ell_3 + \ell_4$, and hence their sum is even and we may rewrite
\begin{align}
i^{-(\ell_1 + \ell_2 + \ell_3 + \ell_4)} = i^{2\times -(\ell_1 + \ell_2 + \ell_3 + \ell_4)/2} = (-1)^{-(\ell_1 + \ell_2 + \ell_3 + \ell_4)/2} = (-1)^{(\ell_1 + \ell_2 + \ell_3 + \ell_4)/2}.
\end{align}
The phase in the last line of equation (\ref{eqn:41}) comes from the isotropic functions' behavior under conjugation (\cite{Cahn} equation 11), which for $N = 4$ gives $\mathcal{P}_{\ell_1 \ell_2 (\ell_{12}) \ell_3 \ell_4}^* = (-1)^{\ell_1 + \ell_2 + \ell_3 + \ell_4} \mathcal{P}_{\ell_1 \ell_2 (\ell_{12}) \ell_3 \ell_4}$. In particular, the harmonics in equation (\ref{eqn:40}) are conjugated and thus give rise to a conjugated isotropic function (\cite{Cahn} equation 10), which we then transform to a non-conjugated one. The phases due to the intermediate angular momentum $\ell_{12}$ and its projective quantum number $m_{12}$ are absorbed in the definition of the isotropic functions.

Now, equating our two different ways of expanding the plane wave, Eqs. (\ref{eqn:lhs_four}) and (\ref{eqn:41}), and eliminating appropriate factors of $4\pi$ on each side, we find
\begin{align}
&H_4 \equiv \sum_{\ell}  (2 \ell + 1) \mathcal{L}_{\ell} (\hat{R} \cdot \hat{R}') j_{\ell}(pR) j_{\ell}(p'R') j_{\ell}(p u_{12}) j_{\ell}(p' u_{12}) = \nonumber\\
&(4\pi)^2 \sum_{\ell_1 \ell_2 \ell_{12}} \sum_{\ell_3 \ell_4}  (-1)^{(\ell_1 + \ell_2 + \ell_3 + \ell_4)/2}  j_{\ell_1}(pr_1) j_{\ell_2}(pr_2) j_{\ell_3} (p'r_3) j_{\ell_4} (p' r_4) j_{\ell_{12}} (p u_{12}) j_{\ell_{12}} (p' u_{12})
\mathcal{D}^{\rm P} \sqrt{2\ell_{12} + 1} \nonumber\\
&\qquad \times (-1)^{\ell_1 +\ell_2 + \ell_3 + \ell_4} \begin{pmatrix}
\ell_1 & \ell_2 & \ell_{12}\\
0 & 0 & 0
\end{pmatrix}
 \begin{pmatrix}
\ell_{12} & \ell_3 & \ell_4\\
0 & 0 & 0
\end{pmatrix}
\mathcal{P}_{\ell_1 \ell_2 (\ell_{12}) \ell_3 {\ell_4}}(\hat{r}_1, \hat{r}_2, \hat{r}_3, \hat{r}_4).
\label{eqn:true_4}
\end{align}
The zero-projective quantum number 3-$j$ symbols imply that the parity of $\ell_{12}$ must equal that of $\ell_1 + \ell_2$ and also that of $\ell_3 + \ell_4$; hence the parity of $\ell_1 + \ell_2 + \ell_3 + \ell_4$ must be even. This sum gives the parity of $\mathcal{P}_{\ell_1 \ell_2 (\ell_{12}) \ell_3 \ell_4}$. This means that, as was the case for $N = 3$ as well, our approach thus far only allows us to explore even-parity isotropic basis functions. Furthermore, the parity of the intermediate must be equal to that of the first two primary angular momenta and of the last two. Within the category of overall even basis functions, this gives a further restriction on what intermediates can be generated. We discuss how to generate odd-parity isotropic basis functions in  \S\ref{sec:odd}.

We now inspect equation (\ref{eqn:true_4}) to confirm that it will give us a way to control both the primary angular momenta and the intermediate. As usual, we must, after expanding each side in a power series (and not necessarily \textit{a priori} restricting $\ell =0$ on the left-hand side) match the overall power of $p$ on each side; here we demand that the power of $p'$ match as well. Then, comparing powers of the $r_i$ on each side and demanding a match fixes the $\ell_i$, up to the need to account for sub-leading terms in the sBf expansions and to subtract them and the basis functions that go with them off, as we did before. The only addition beyond previous sections is that now, the powers of $u_{12}$ may also be matched on each side, and this enables controlling the intermediate, $\ell_{12}$. We now show two examples of how this proceeds in practice.

\subsection{$N = 4$ Examples}
\subsubsection{$\mathcal{P}_{10 (1) 01}$}
Here, looking at equation (\ref{eqn:true_4}), we have $\ell_1 = \ell_4 = 1$, and this means we need combinations on the left- and right-hand sides that behave as $r_1 r_4$. We also observe that, since $\ell_{12}  = 1$, to match the powers of $u_{12}$ on the right-hand side that we will have from expanding the two $j_1$'s to leading order, we need $u_{12}^2$ on the left-hand side. This implies $\ell = 1$ on the left-hand side, if we look at the leading order of the sBf series that would result. In particular, since the series for $j_0$ is even parity, the sub-leading terms if we set $\ell = 0$ on the left-hand side would give $u_{12}^4$ and higher powers, so $\ell = 0$ cannot be relevant.

Setting $\ell = 1$, we thus have the left-hand side of equation (\ref{eqn:true_4}) as
\begin{align}
    H_{4, 10 (1) 01} \to 3 \hat{R}\cdot\hat{R}'\, j_1(pR) j_1(p'R') j_1(p u_{12}) j_1(p' u_{12}),
    \label{eqn:H4-10101}
\end{align}
where our additional subscript on $H_4$ indicates that it is just the terms in the sum of equation (\ref{eqn:true_4}) relevant for $\mathcal{P}_{10 (1) 01}$. We adopt this notation here for clarity now that we have an intermediate as well, rather than the vertical bar and subscripted $r_i$ of previous sections. In the vertical bar notation, we would have $H_{4}\big |_{r_1 r_4 u_{12}^2}$.

Now expanding each sBf to leading order, we find from equation (\ref{eqn:H4-10101}) that
\begin{align}
    &H_{4, 10 (1) 01} \to 
    3 \hat{R}\cdot\hat{R}'\, \frac{pR}{3} \,\frac{p'R'}{3} \,\frac{p u_{12}}{3} \,\frac{p' u_{12}}{3}
    = \frac{1}{27}\, {\bf R} \cdot {\bf R'}\, p^2 p'^2 u_{12}^2.
\end{align}
Next, we expand the dot product of ${\bf R} \cdot {\bf R'}$ and retain only the term in $r_1 r_4$. We find that this term has coefficient unity, so we have
\begin{align}
    &H_{4, 10 (1) 01} \to 
    3 \hat{R}\cdot\hat{R}'\, \frac{pR}{3} \frac{p'R'}{3} \frac{p u_{12}}{3} \frac{p' u_{12}}{3}
    = \frac{1}{27}\, r_1 r_4\, (\hat{r}_1 \cdot{\hat{r}_4}) p^2 p'^2 u_{12}^2.
\end{align}
We now set this result equal to the right-hand side of equation (\ref{eqn:true_4}). We first expand each sBf in a series, and note that demanding equality of the powers of $p$, $p'$, $u_{12}$, and the $r_i$ removes the sums over all $\ell_i$ and $\ell_{12}$. We then cancel the powers $p$, $p'$, $u_{12}$, and the $r_i$, and evaluate the factor $\mathcal{D}^{\rm P}$, $\sqrt{2 \ell_{12} + 1}$, and the 3-$j$ symbols. We find
\begin{align}
    \frac{1}{27}\, \hat{r}_1 \cdot \hat{r}_4 = (4\pi)^2 (-1) \frac{1}{3\cdot 27} \sqrt{3}^3 \left(-\frac{1}{\sqrt{3}} \right)^2 \mathcal{P}_{10(1)01}.
\end{align}
Simplifying and solving for $\mathcal{P}_{10(1)01}$, we obtain
\begin{align}
    \mathcal{P}_{10(1)01}(\hat{r}_1, \hat{r}_2, \hat{r}_3, \hat{r}_4) = -\frac{\sqrt{3}}{(4\pi)^2}\, \hat{r}_1 \cdot \hat{r}_4,
\end{align}
in agreement with \cite{Cahn} Appendix A.3., second line.

\subsubsection{$\mathcal{P}_{11 (0) 11}$}
Here, we use the same approach as in the previous subsection. Now we see that on the right-hand side of equation (\ref{eqn:true_4}), we will need $p^2 p'^2 u_{12}^0$ and combinations behaving as $r_1 r_2 r_3 r_4$. We might at first think $\ell = 1$ on the left-hand side is the correct choice, as this gives the right powers of $p$ and $p'$ if the sBfs there are expanded to leading order. However, this would give the wrong power of $u_{12}$; we would have $u_{12}^2$ at leading order, but we need $u_{12}^0$. If we set $\ell = 0$ and take the the second-order terms from expanding $j_0(pR) j_0(p'R')$ but the leading-order terms from expanding $j_0(pu_{12}) j_0(p'u'_{12})$, we will then have the desired powers of $p$, $p'$, and $u_{12}$. We have
\begin{align}
    H_{4, 11 (0) 11} \to \frac{1}{36} (pR)^2 (p' R')^2. 
\end{align}
In the vertical bar notation, we would have $H_{4}\big |_{r_1 r_2 r_3 r_4}$.
We now expand the powers of $R$ and $R'$ and retain only the term in $r_1 r_2 r_3 r_4$, which has a coefficient of $4$. We find
\begin{align}
    H_{4, 11 (0) 11} \to \frac{1}{9} p^2 p'^2 r_1 r_2 r_3 r_4\, (\hat{r}_1 \cdot \hat{r}_2) (\hat{r}_3 \cdot \hat{r}_4). 
\end{align}
Returning to equation (\ref{eqn:true_4}) and demanding equality of all powers, we see this eliminates the sums over the $\ell_i$ and over $\ell_{12}$, and we have, cancelling the factor of $p^2 p'^2 r_1 r_2 r_3$, that
\begin{align}
    \frac{1}{9}\, (\hat{r}_1 \cdot \hat{r}_2) (\hat{r}_3 \cdot \hat{r}_4) = (4\pi)^2 \frac{1}{3^4} \sqrt{3}^4 \left(-\frac{1}{\sqrt{3}} \right)^2 \mathcal{P}_{11 (0) 11} (\hat{r}_1, \hat{r}_2, \hat{r}_3, \hat{r}_4). 
\end{align}
Solving for $\mathcal{P}_{11 (0) 11}$ and simplifyng, we obtain
\begin{align}
    \mathcal{P}_{11 (0) 11}(\hat{r}_1, \hat{r}_2, \hat{r}_3, \hat{r}_4) = \frac{3}{(4\pi)^2}\, (\hat{r}_1 \cdot \hat{r}_2) (\hat{r}_3 \cdot \hat{r}_4),
\end{align}
in agreement with \cite{Cahn} Appendix A.3., first line.

\section{Generating Function for $N=5$}
\label{sec:N5}

Let us now apply the same approach to deriving the generating function for the $N = 5$ isotropic basis functions. We consider
\begin{align}
P_5 \equiv e^{i \vec{p} \cdot (\vec{r}_1 + \vec{r}_2)} e^{i \vec{p}' \cdot \vec{r}_3} e^{i \vec{p}'' \cdot ( \vec{r}_4 +\vec{r}_5)} e^{i \vec{p} \cdot \vec{u}_{12}}
e^{i \vec{p}' \cdot \vec{u}_{12}} e^{i \vec{p}' \cdot \vec{u}_{123}} e^{i \vec{p}'' \cdot \vec{u}_{123}}.
\end{align}
Defining $R \equiv |\vec{r}_1 + \vec{r}_2|$, $R' \equiv |\vec{r}_3|$, and $R'' \equiv |\vec{r}_4 + \vec{r}_5|$, we have
\begin{align}
\label{eqn:N5st}
P_5 = &(4\pi)^7 \sum_{\ell \ell' \ell''} \sum_{\ell_{12} \ell_{12}'} \sum_{\ell_{123}' \ell_{123}''} \sum_{{\rm all \; m}} i^{\ell + \ell' + \ell'' + \ell_{12} + \ell_{12}' + \ell_{123}' + \ell_{123}''} j_{\ell} (pR) j_{\ell'} (p'R') j_{\ell''} (p'' R'')  j_{\ell_{12}}( p u_{12})  j_{\ell_{12}'} (p' u_{12}) \nonumber\\
& \times j_{\ell_{123}'}(p' u_{123}) j_{\ell_{123}''} (p'' u_{123})  Y_{\ell m} (\hat{R})
Y_{\ell m}^*(\hat{p})
Y_{\ell' m'}(\hat{R}')
Y_{\ell' m'}^*(\hat{p}')
Y_{\ell'' m''}(\hat{R}'') Y_{\ell'' m''}^*(\hat{p}'')\\
&\times Y_{\ell_{12} m_{12}}(\hat{u}_{12}) 
Y_{\ell_{12} m_{12}}^*(\hat{p}) Y_{\ell_{12}' m_{12}'}(\hat{u}_{12}) Y_{\ell_{12}' m_{12}'}^*(\hat{p}')
Y_{\ell_{123}' m_{123}'}(\hat{u}_{123})
Y_{\ell_{123}' m_{123}'}^*(\hat{p}')\nonumber\\
& \times Y_{\ell_{123}'' m_{123}''}(\hat{u}_{123})
Y_{\ell_{123}'' m_{123}''}^*(\hat{p}'').\nonumber
\end{align}
Now performing an integral over $d \Omega_p d\Omega_{p'} d\Omega_{p''}$
we obtain
\begin{align}
\label{eqn:p5-pre-simp}
P_5 &= (4\pi)^7 \sum_{\rm all} i^{(\cdots)} j_{\ell}(pR) \cdots j_{\ell_{123}''} (p'' u_{123})   Y_{\ell m}(\hat{R}) Y_{\ell' m'}(\hat{R}') Y_{\ell_{12} m_{12}}(\hat{u}_{12}) Y_{\ell_{12}' m_{12}'}(\hat{u}_{12})  \\
&\times Y_{\ell_{123}' m_{123}'}(\hat{u}_{123}) Y_{\ell_{123}'' m_{123}''}(\hat{u}_{123}) Y_{\ell'' m''}(\hat{R}'') \delta^{\rm K}_{\ell \ell_{12}} (-1)^{m_{12}} \delta^{\rm K}_{m - m_{12}} \mathcal{G}_{\ell' \ell_{12}' \ell_{123}'}^{m' m_{12}' m_{123}'}  \delta^{\rm K}_{\ell'' \ell_{123}''}  (-1)^{m_{123}''}\delta^{\rm K}_{m'' -m_{123}''},\nonumber
\end{align}
where the subscript ``${\rm all}$'' on the sum indicates to sum over all the subscripts on the right-hand side, and the $\cdots$ in the phase indicate the same phase as in equation (\ref{eqn:N5st}). The ellipsis $(\cdots)$ in between the sBfs indicate they are all copied over from equation (\ref{eqn:N5st}). We note that there were just two harmonics involving $\hat{p}$ and also just two involving $\hat{p''}$, hence the Kronecker deltas for those angular momenta, as well as the phases and flipped signs on the projective quantum numbers due to using the conjugate identity. We also used that the Gaunt integral is real, so taking its conjugate, which is equivalent to integrating over three conjugated spherical harmonics, leaves it unchanged.

Simplifying equation (\ref{eqn:p5-pre-simp}), we obtain
\begin{align}
P_5 &= (4\pi)^7 \sum_{\rm all} i^{2 (\ell + \ell'') + \ell' + \ell_{12}' + \ell_{123}'} 
 j_{\ell}(pR) \cdots j_{\ell''} (p'' u_{123})
Y_{\ell m}(\hat{R})
Y_{\ell' m'}(\hat{R}')
Y_{\ell'' m''}(\hat{R}'')
\mathcal{G}_{\ell' \ell_{12}' \ell_{123}'}^{m' m_{12}' m_{123}'} \nonumber\\
&\qquad \times Y_{\ell m}^* (\hat{u}_{12})
Y_{\ell_{12}' m_{12}'} (\hat{u}_{12})
Y_{\ell_{123}' m_{123}'}(\hat{u}_{123})
Y_{\ell'' m''}^* (\hat{u}_{123}).
\end{align}
We now further integrate over $d \Omega_{12} d\Omega_{123}$ and obtain
\begin{align}
P_5 &= (4\pi)^7 \sum_{\rm all} (-1)^{\ell + \ell'}  i^{\ell + \ell' + \ell''}   
j_{\ell} (pR) j_{\ell'} (p'R') j_{\ell''} (p'' R'') \nonumber\\
&\times j_{\ell}( p u_{12})  j_{\ell} (p' u_{12})
j_{\ell''} (p' u_{123})
j_{\ell''}(p'' u_{123})
Y_{\ell m}(\hat{R})
Y_{\ell' m'}(\hat{R}')
Y_{\ell'' m''}(\hat{R}'')
\mathcal{G}_{\ell'  \ell \ell''}^{m' m m''}.
\label{eqn:lhs_N_5}
\end{align} 
We note that the Gaunt integral above contains a zero-projective quantum number 3-$j$ symbol, which in turn guarantees that $\ell + \ell' + \ell''$ is even and hence that $i$ to this power is real.  For this reason we may also reorder the Gaunt to be in order of ascending primes without picking up any overall phase. We observe that the generating function for the $N = 5$ isotropic basis functions thus turns out to be a tripolar spherical harmonic in $\hat{R},\, \hat{R}',$ and $\hat{R}''$. It now remains to obtain the right-hand side of the expansion of the plane wave here, which will explicitly display the relationship of the above to the $N = 5$ basis functions. This is extensive but conceptually much simpler than was the expansion for the left-hand side. We simply write
\begin{align}
\label{eqn:big_rhs_5}
P_5 &=  e^{i \vec{p} \cdot \vec{r}_1}
e^{i \vec{p} \cdot \vec{r}_2}
e^{i \vec{p}' \cdot \vec{r}_3}
e^{i \vec{p}'' \cdot \vec{r}_4}
e^{i \vec{p}'' \cdot \vec{r}_5}
e^{i \vec{p}' \cdot \vec{u}_{12}}
e^{i \vec{p} \cdot \vec{u}_{12}}
e^{i \vec{p}' \cdot \vec{u}_{123}}
e^{i \vec{p}'' \cdot \vec{u}_{123}} = \nonumber\\
&(4\pi)^9 \sum_{\rm all} i^{\ell_1 + \ell_2 + \ell_3 + \ell_4 + \ell_5 + \ell_{12} + \ell_{12}' + \ell_{123}' + \ell_{123}''}\nonumber\\
&\times j_{\ell_1}(pr_1) j_{\ell_2}(pr_2) j_{\ell_3}(p' r_3) j_{\ell_4}(p'' r_4) j_{\ell_5}(p'' r_5) j_{\ell_{12}}(p u_{12}) j_{\ell'_{12}}(p' u_{12}) j_{\ell_{123}'}(p' u_{123}) j_{\ell_{123}''} (p'' u_{123})\nonumber\\
&\times \mathcal{Y}_{\ell_1 \ell_2 \ell_{12}}^{m_1 m_2 m_{12} *} (\hat{p})
\mathcal{Y}_{\ell_3 \ell'_{12} \ell_{123}'}^{m_3 m_{12}' m_{123}' *} (\hat{p}')
\mathcal{Y}_{\ell_4 \ell_5 \ell_{123}''}^{m_4 m_5 m_{123}'' *} (\hat{p}'')
Y_{\ell_1 m_1}(\hat{r}_1)
Y_{\ell_2 m_2}(\hat{r}_2)
Y_{\ell_3 m_3}(\hat{r}_3)
Y_{\ell_4 m_4}(\hat{r}_4)
Y_{\ell_5 m_5}(\hat{r}_5) \nonumber\\
& \times Y_{\ell_{12} m_{12}} (\hat{u}_{12}) Y_{\ell_{12}' m_{12}'} (\hat{u}_{12}) 
Y_{\ell_{123}' m_{123}'} (\hat{u}_{123}) Y_{\ell_{123}'' m_{123}''} (\hat{u}_{123}),
\end{align}
with 
\begin{align}
\mathcal{Y}_{a b c}^{d e f *}(\hat{x}) \equiv Y_{a d}^*(\hat{x}) Y_{b e}^*(\hat{x}) Y_{c f}^*(\hat{x}).
\end{align}
Now integrating equation (\ref{eqn:big_rhs_5}) over $d\Omega_p d\Omega_{p'} d\Omega_{p''}$ removes the three $\mathcal{Y}$ in favor of three Gaunt integrals, as
\begin{align}
&\int d\Omega_p d\Omega_{p'} d\Omega_{p''} \,\mathcal{Y}_{\ell_1 \ell_2 \ell_{12}}^{m_1 m_2 m_{12} *} (\hat{p})
\mathcal{Y}_{\ell_3 \ell'_{12} \ell_{123}'}^{m_3 m_{12}' m_{123}' *} (\hat{p}')
\mathcal{Y}_{\ell_4 \ell_5 \ell_{123}''}^{m_4 m_5 m_{123}'' *} (\hat{p}'')  \\
&\qquad = \mathcal{G}_{\ell_1 \ell_2 \ell_{12}}^{m_1 m_2 m_{12}} 
\mathcal{G}_{\ell_3 \ell_{12}' \ell_{123}'}^{m_3 m_{12}' m_{123}'} 
\mathcal{G}_{\ell_4 \ell_5 \ell_{123}''}^{m_4 m_5 m_{123}''},\nonumber
\end{align}
where the second and third Gaunt integrals above may be non-cyclically permuted without acquiring any net phase, as the phases from permutation will vanish because each involves an even power of $(-1)$, as guaranteed by the presence of a zero-projective quantum number 3-$j$ symbol within each Gaunt. 

We now integrate over $d \Omega_{12} d\Omega_{123}$, finding that 
\begin{align}
&\int d \Omega_{12} d\Omega_{123}\, Y_{\ell_{12} m_{12}} (\hat{u}_{12}) Y_{\ell_{12}' m_{12}'} (\hat{u}_{12}) 
Y_{\ell_{123}' m_{123}'} (\hat{u}_{123}) Y_{\ell_{123}'' m_{123}''} (\hat{u}_{123})  \nonumber\\
&\qquad = \delta^{\rm K}_{\ell_{12}' \ell_{12}} (-1)^{m_{12}} \delta^{\rm K}_{m_{12}' -m_{12}}
\delta^{\rm K}_{\ell_{123}'' \ell_{123}'}
(-1)^{m_{123}'} \delta^{\rm K}_{m_{123}'' -m_{123}'}.
\end{align} 
Putting the results above together and rewriting in terms of the $N = 5$ isotropic basis functions using their definition, equation (\ref{eqn:iso_5}), we find that
\begin{align}
P_5 &= (4\pi)^9 \sum_{\rm all} (-1)^{(\ell_1 + \ell_2  + \ell_3 + \ell_4 + \ell_5)/2} 
j_{\ell_1}(pr_1) j_{\ell_2}(pr_2) j_{\ell_3}(p' r_3) j_{\ell_4}(p'' r_4) j_{\ell_5}(p'' r_5) j_{\ell_{12}}(p u_{12}) j_{\ell_{12}}(p' u_{12})\nonumber\\
&\times j_{\ell_{123}}(p' u_{123}) j_{\ell_{123}} (p'' u_{123})
\begin{pmatrix}
\ell_1 & \ell_2 & \ell_{12}\\
0 & 0 & 0
\end{pmatrix}
\begin{pmatrix}
\ell_{12} & \ell_3 & \ell_{123}\\
0 & 0 & 0
\end{pmatrix}
\begin{pmatrix}
\ell_{123} & \ell_4 & \ell_5\\
0 & 0 & 0
\end{pmatrix}
\mathcal{D}^{\rm P} \nonumber\\
&\times\sqrt{ 2\ell_{12} + 1} \sqrt{2 \ell_{123} + 1}  \,\mathcal{P}_{\ell_1 \ell_2 (\ell_{12}) \ell_3 (\ell_{123}) \ell_4 \ell_5}(\hat{r}_1, \hat{r}_2, \hat{r}_3, \hat{r}_4, \hat{r}_5).
\label{eqn:rhs_N5}
\end{align}
We note that as a last step we have renamed $\ell_{123}'$ as $\ell_{123}$ without loss of generality and to be consistent with the convention in \cite{Cahn}. By comparing the equation above with equation (\ref{eqn:lhs_N_5}) we see that we have obtained the generating function for the $N = 5$ isotropic basis functions. We also see that expanding the above in small $p$, $p'$, and $p''$ allows us to select a total of the primary $\ell_i$, and then the power of each $r_i$ to control the individual $\ell_i$. Then, the powers of respectively $u_{12}$ and $u_{123}$ may be used to control the intermediate angular momenta $\ell_{12}$ and $\ell_{123}$.

\section{Generating Function for Odd-Parity Isotropic Functions}
\label{sec:odd}
As noted earlier, the generating functions thus far presented produce only even isotropic basis functions. However, in 3D, one can have odd parity basis functions of three or more arguments, \textit{i.e.} for $N \geq 3$. Below we outline how the generating function may be obtained for $N=3$; we defer the $N > 3$ cases to future work. The $N=3$ odd-parity basis functions are already sufficient to perform a parity violation search with the galaxy 4-Point Correlation Function, as first proposed in \cite{parity-prl} and carried out in \cite{hou-parity}.

\subsection{For $N = 3$}
We notice that $\mathcal{P}_{111}$ is the simplest odd-parity function (recall there are none for $N = 2$). We consider if we can use it as a building block to produce the other odd parity $N = 3$ functions. Let us consider its product with an arbitrary even-parity $N = 3$ function.

The linearization identity for two isotropic functions of the same arguments is 
\begin{align}
\mathcal{P}_{\Lambda} \mathcal{P}_{\Lambda'} = \sum_{\Lambda''} \mathcal{G}_{\Lambda \Lambda' \Lambda''} \mathcal{P}_{\Lambda''},
\label{eqn:lin_odd}
\end{align}
where $\mathcal{G}$ is a generalized Gaunt integral as defined in \cite{Cahn}; it is explicitly given for $N=3$ in their equation (62). We set $\Lambda = (\ell_1, \ell_2, \ell_3)$ and $\Lambda' = (1, 1, 1)$, since using the lowest-order odd isotropic basis function will convert even functions into their nearest-lying odd neighbors. We can view $\mathcal{
P
}_{111}$ almost as an operator, which raises/lowers an even state to the proximate odd states. We find
\begin{align}
\mathcal{G}_{\Lambda \Lambda' \Lambda''}  = (4\pi/3)^{-3/2} \bigg[ \prod_{i = 1}^3 (2\ell_i + 1) 
\begin{pmatrix}
\ell_i & 1 & \ell_i''\\
0 & 0 & 0
\end{pmatrix}\bigg]
\left\{\begin{matrix}
\ell_1 & 1 & \ell_1'' \\
\ell_2 & 1 & \ell_2'' \\
\ell_3 & 1 & \ell_3'' \\
\end{matrix}\right\}.
\end{align}
The $3 \times 3$ matrix is a Wigner 9-$j$ symbol. There will be two allowed values of each $\ell''_i$, namely $\ell''_i = \ell_1 + 1, \ell_1 - 1$; thus the right-hand side of equation (\ref{eqn:lin_odd}) will contain eight odd-parity isotropic functions. The orthogonality relation for 9-$j$ symbols, NIST DLMF 34.7.2,\footnote{\url{https://dlmf.nist.gov/34.7} We note that we can use the reflection symmetry of the 9-$j$ to bring it to the form required by the orthogonality relation, where the two non-matching $j$'s are the bottom left two elements. Then orthogonality will work on the coefficients where the bottom right element is matched, yet some symbols will not have that as the case and they will give the ``anomalous'' contribution.} shows that if we sum over two of the $j$'s then we can pick out a desired isotropic function plus an anomalous contribution that involves other isotropic functions. Doing this with all the possible choices of the two $j$'s to sum over will result in eight independent linear equations for the eight odd isotropic functions; these may then be solved. This will give an expression for each of them in terms of the products $\mathcal{P}_{\ell_1 \ell_2 \ell_3} \mathcal{P}_{1 1 1}$ weighted by angular momentum symbols; replacing $\mathcal{P}_{\ell_1 \ell_2 \ell_3}$ by its Cartesian form, which can be obtained from the foregoing sections of the present work, then gives a Cartesian form for the desired odd isotropic functions.

\section{Expansion of the Dirac Delta Function}
\label{sec:dirac}
We now outline how to obtain quickly the expansion of the Dirac Delta function of multiple arguments in terms of the appropriate isotropic basis functions. We recall that the Delta function has spherical symmetry and is also the 3D inverse FT of unity; hence
\begin{align}
\delta_{\rm D}^{[3]}(R) = \int \frac{p^2 dp}{2\pi^2} \; j_0(pR).
\end{align}
$\delta_{\rm D}^{[3]}$ is a 3D Dirac Delta function, and we note it is spherically symmetric and hence depends only on $R$, the magnitude of $\bf{R}$. This may be seen easily by recalling that a 3D Delta function is just the zero-width limit of a 3D Gaussian, which is spherically symmetric. $p$ is a vector magnitude and so is positive semi-definite; thus, here and throughout, unless otherwise indicated any integral over $p$ ranges from zero to infinity and we will not write the bounds explicitly.

The integral above may be shown using the plane wave expansion into spherical harmonics and spherical Bessel functions along with the spherical symmetry of the Delta function's Fourier-space counterpart, unity. Thus, if we integrate both sides of our generating function relation against $p^2 dp/(2\pi^2)$, we obtain immediately the expansion of the Delta function in the isotropic basis. Schematically, it is
\begin{align}
\delta_{\rm D}^{[3]} ( R ) \to \sum_{{\rm all} \; \ell} \int \frac{p^2 dp}{2\pi^2}\, j_{\ell_1}(pr_1 ) \cdots j_{\ell_N} (pr_N) \mathcal{P}^{(N)}_{\ell_1 \cdots \ell_N},
\end{align}
where $R \equiv |\sum_i^N {\bf r}_i|$ (we recall equation \ref{eqn:R_Def}) and the arguments of the $N$-argument isotropic basis function $\mathcal{P}^{N}_{\ell_1 \cdots \ell_N}$ are the $\hat{r}_i$.

\subsection{Delta Function for $N=2$ and $N=3$}
This is straightforward for $N = 2$ and $N = 3$, when there are no intermediate angular momenta; we simply write down the results below. 
For $N=2$, we have
\begin{align}
\int \frac{p^2 dp}{2\pi^2}\, j_0(pR) = (4\pi)^{3/2} \sum_{\ell} \sqrt{2\ell + 1}\; \mathcal{P}_{\ell \ell}(\hat{r}_1, \hat{r}_2) \int \frac{p^2 dp}{2\pi^2}\, j_{\ell}(pr_1) j_{\ell}(pr_2),
\end{align}
and the overlap integral on the right-hand side may be simplified to $1/(4\pi r_1 r_2)\, \delta_{\rm D}^{[1]}(r_1 - r_2)$ if desired.

For $N=3$, we have 
\begin{align}
\label{eqn:N3_delta}
&\int \frac{p^2 dp}{2\pi^2}\, j_0(pR) = \\
& (4\pi)^2 \sum_{\ell_1 \ell_2 \ell_3} (-1)^{(\ell_1 +\ell_2 +\ell_3)/2} \mathcal{C}_{\ell_1\ell_2 \ell_3} 
\begin{pmatrix}
\ell_1 & \ell_2 & \ell_3\\
0 & 0 & 0
\end{pmatrix}
 \mathcal{P}_{\ell_1 \ell_2 \ell_3} (\hat{r}_1, \hat{r}_2, \hat{r}_3) \int \frac{p^2 dp}{2\pi^2}\, j_{\ell_1}(pr_1) j_{\ell_2} (pr_2) j_{\ell_3}(pr_3).\nonumber
\end{align}
The sum of the $\ell_i$ is guaranteed to be even by the zero-projective quantum number 3-$j$ symbol, thus our simplification of the phase we obtained from the plane wave expansions. 

The triple-sBf integral on the right-hand side can be done using a result in \cite{mehrem_2_91}, where they denote it $I$ (more precisely our integral above is  $I/(2\pi^2)$ under their definition for $I$). We have 
\begin{align}
&I(\ell_1, \ell_2, \ell_3; r_1, r_2, r_3) = \frac{\pi  \beta(\Delta)}{4 r_1 r_2 r_3} i^{\ell_1 + \ell_2 + \ell_3} \sqrt{2\ell_3 + 1} \left( \frac{r_1}{r_3}\right)^{\ell_3}
\begin{pmatrix}
\ell_1 & \ell_2 & \ell_3\\
0 & 0 & 0
\end{pmatrix}^{-1}
\sum_{\mathcal{L}}^{\ell_3} {{2\ell_3} \choose {2 \mathcal{L}}}^{1/2} \left( \frac{r_2}{r_1}\right)^{\mathcal{L}} \nonumber\\
&\times \sum_{\ell} (2\ell + 1) 
\begin{pmatrix}
\ell_1 & \ell_3-\mathcal{L} & \ell\\
0 & 0 & 0
\end{pmatrix}
\begin{pmatrix}
\ell_2 & \mathcal{L} & \ell\\
0 & 0 & 0
\end{pmatrix}
\left\{\begin{matrix}
\ell_1 & \ell_2 & \ell_3\\
\mathcal{L} & \ell_3 - \mathcal{L} & \ell
\end{matrix}\right\}
\mathcal{L}_{\ell}(\Delta),
\end{align}
with 
\begin{align}
&\Delta \equiv \frac{r_1^2 + r_2^2 - r_3^2}{2 r_1 r_2},\;\;\; \beta(\Delta) \equiv \theta(1 - \Delta) \theta(1 + \Delta),\nonumber\\
& \theta(y) = 0,\; y<0;\;\;\; 1/2,\; y = 0;\;\;\; 1,\; y>0.
\end{align}
We note that the Legendre polynomial on the right-hand side does not explicitly depend upon the individual unit vectors $\hat{r}_1, \hat{r}_2,$  or $\hat{r}_3$; its argument is actually just the cosine of the angle between $r_1$ and $r_2$ in a triangle with sides $r_1, r_2$, and $r_3$. We also note that the $2 \times 3$ matrix in curly brackets is a Wigner 6-$j$ symbol (\textit{e.g.} NIST DLMF 34.4\footnote{\url{https://dlmf.nist.gov/34.4}}), defined as a sum over projective quantum numbers of a specified product of 3-$j$ symbols. 

\subsection{Delta Function for $N=4$: Naive Approach}
\label{subsec:delta-naive}
For $N = 4$ and $N =5$, the treatment of the intermediates requires additional consideration. However, an interesting observation is that, for the Delta function derivation, we may begin with the ``naive'' (incorrect) $N=4$ generating function equation (\ref{eqn:naive_4}) and still obtain a correct expansion. We first do this, and then begin with the proper generating function, showing that, interestingly, using these two approaches enables expanding slightly different versions of the Delta function. 
Proceeding first from equation (\ref{eqn:naive_4}), we have, with $\vec{R} \equiv \vec{r}_1  + \vec{r}_2 + \vec{r}_3 + \vec{r}_4$ and $R \equiv |\vec{R}|$ there and here,
\begin{align}
&\int \frac{p^2 dp}{2\pi^2} \, j_0(pR) = (4\pi)^2 \sum_{\ell_1, \ldots, \ell_4} (-1)^{(\ell_1 + \ell_2 + \ell_3 + \ell_4)/2}\, \mathcal{D}^{\rm P} \sum_{\ell_{12}} \sqrt{2\ell_{12} + 1} 
\begin{pmatrix}
\ell_1 & \ell_2 & \ell_{12} \\
0 & 0 & 0
\end{pmatrix}
\begin{pmatrix}
\ell_{12} & \ell_3 & \ell_4 \\
0 & 0 & 0
\end{pmatrix}\nonumber\\
&\times (-1)^{\ell_{12}}
\mathcal{P}_{\ell_1 \ell_2 (\ell_{12}) \ell_3 \ell_4}(\hat{r}_1, \hat{r}_2, \hat{r}_3, \hat{r}_4)
\int \frac{p^2 dp}{2\pi^2} \, j_{\ell_1}(pr_1)  j_{\ell_2}(pr_2)  j_{\ell_3}(pr_3)  j_{\ell_4}(pr_4).
\end{align}
We simplified the phase since the two 3-$j$ symbols imply that the sum of the primary $\ell_i$ is even. If we denote the quadruple-sBf overlap integral on the right-hand side (including the factor of $1/(2\pi^2)$) as $I_4$, it may be evaluated using a technique presented by \cite{mehrem_pw_pub}, \S6. One rewrites the integral as a double integral where the outer integrand is a pair of triple-sBf integrals; this approach relies on the fact that a double-sBf overlap integral with the appropriate weight and at matching orders gives a Dirac Delta function, as
\begin{align}
    \int x^2 dx\; j_{\ell}(ax)
    j_{\ell}(bx) = \frac{\pi}{2 ab} \,\delta_{\rm D}^{[1]}(a-b),
\end{align}
\textit{e.g.} NIST DLMF 1.17.14.\footnote{\url{https://dlmf.nist.gov/1.17}}

We have
\begin{align}
I_4 = \frac{2}{\pi} \cdot \frac{1}{2\pi^2}  \int q^2 dq \left[ \int p^2 dp\; j_{\ell_1}(pr_1) j_{\ell_2}(pr_2) j_{L}(q p) \int p'^2 dp' \; j_{\ell_3}(p' r_3) j_{\ell_4}(p' r_4) j_L(q p') \right].
\end{align}
The question now becomes whether we can always evaluate the two triple-sBf integrals required above. Examining the result of \cite{mehrem_2_91} for each of the three-sBf integrals $I$ we require, we see that we may as long as $L$ is chosen so as to be in a triangular relation with both $\ell_1, \ell_2$ and $\ell_3, \ell_4$. If we choose that $L \equiv \ell_{12}$, this will always be guaranteed, and indeed provides an interesting connection to the idea of coupling two angular momenta into an intermediate; this now happens on the orders of the sBfs. Making this choice, we find that
\begin{align}
I_4 = \frac{2}{\pi} \cdot \frac{1}{2\pi^2} \int q^2 dq \left[ I(\ell_1, \ell_2, \ell_{12}; r_1, r_2, q)\; I(\ell_{12}, \ell_3, \ell_4; q, r_3, r_4) \right] \equiv \pi^{-3} \,\bar{I}_4.
\end{align}
Hence, the Dirac Delta function may be expanded in the isotropic basis as
\begin{align}
    &\delta_{\rm D}^{[3]}(|{\bf r}_1 + {\bf r}_2 +{\bf r}_3 + {\bf r}_4| ) = \sum_{\ell_i} c_{\ell_1 \ell_2 \ell_3 \ell_4}(r_1, r_2, r_3, r_4) \nonumber\\
    &\qquad\qquad \times \sum_{\ell_{12}} (-1)^{\ell_{12}}  \sqrt{2\ell_{12} + 1} 
\begin{pmatrix}
\ell_1 & \ell_2 & \ell_{12} \\
0 & 0 & 0
\end{pmatrix}
\begin{pmatrix}
\ell_{12} & \ell_3 & \ell_4 \\
0 & 0 & 0
\end{pmatrix}
\mathcal{P}_{\ell_1 \ell_2 (\ell_{12}) \ell_3 \ell_4}(\hat{r}_1, \hat{r}_2, \hat{r}_3, \hat{r}_4)
\label{eqn:delta_exp_1}
\end{align}
with 
\begin{align}
    c_{\ell_1 \ell_2 \ell_3 \ell_4}(r_1, r_2, r_3, r_4)
    \equiv
    \frac{16}{\pi}\,
    (-1)^{(\ell_1+ \ell_2 +\ell_3+ \ell_4)/2} \,\mathcal{D}^{\rm P}\,
    \bar{I}_4(r_1, r_2, r_3, r_4), 
    \label{eqn:c-delta-1}
\end{align}
where $\mathcal{D
}^{\rm P}$ is a product of $\sqrt{2\ell_i + 1}$ over all primary angular momenta.

It is notable that the radial expansion coefficients do not depend on the intermediate angular momentum, $\ell_{12}$, which is summed over at fixed $\ell_i$ in our expansion. The symmetry of the Delta function with respect to its four vector arguments $\bf{r}_i$ suggests that any expansion for it should be ``symmetrized'', or summed, over all possible couplings of the four primary angular momenta that pertain to them, and this is indeed what occurs. This summation also means that we may define ``recoupling-symmetrized'' isotropic basis functions by the sum in the second line of equation (\ref{eqn:delta_exp_1}); these are a subset of the full isotropic basis which is suitable for expanding functions that have global interchange symmetry among all arguments, as the Dirac Delta function does. Ultimately, the irrelevance of the intermediate is why we can obtain a correct expansion of the Delta function even using the naive $N = 4$ generating function approach (which does not allow control of the intermediate).

\subsection{Delta Function for $N=4$: Approach with the Correct Generating Function}
We now show that we obtain the same result as in the previous section if we begin with the correct generating function at $N = 4$, which is given by equation (\ref{eqn:true_4}).

Here, it is informative to begin with the Dirac delta function itself. We first write the desired 3D Dirac delta function as the inverse Fourier Transform of unity:
\begin{align}
    \delta_{\rm D}^{[3]} ({\bf R} + {\bf R'}) = \int \frac{d^3 \vec{p}}{(2\pi)^3}\; e^{-i \vec{p} \cdot \vec{R}} e^{-i \vec{p} \cdot \vec{R}'}.
\end{align}
Expanding the plane waves into spherical harmonics and spherical Bessel functions, integrating over $d\Omega_p$, invoking orthogonality, and summing over projective quantum numbers, we obtain
\begin{align}
\delta_{\rm D}^{[3]} ({\bf R} + {\bf R'}) = 
      \sum_{\ell} (-1)^{\ell} (2\ell + 1) \mathcal{L}_{\ell}(\hat{R} \cdot \hat{R}')  \int \frac{p^2 dp}{2\pi^2}\; j_{\ell}(p R) j_{\ell}(p R'),  
      \label{eqn:seventyeight}
\end{align}
If we integrate equation (\ref{eqn:true_4}) over $u_{12}^2 du_{12}$ we will obtain a result resembling our desired $p$ integrand in equation (\ref{eqn:seventyeight}). In particular, we get from the $u_{12}$ integration:
\begin{align}
    \int u_{12}^2 du_{12}\; H_4 = 
    \sum_{\ell} (-1)^{\ell} (2\ell + 1) \mathcal{L}_{\ell}(\hat{R} \cdot \hat{R}') \, j_{\ell}(p R) j_{\ell}(p' R')\,\frac{\pi}{2 p^2} \delta_{\rm D}^{[1]}(p - p'). 
\end{align}
The Delta function will set $p = p'$, so if we integrate over $p$ we can come close to equation (\ref{eqn:seventyeight}); however we do not have the right pre-factor. We thus need to multiply by $p^4/\pi^3$ before integrating. We then have
\begin{align}
    \pi^{-3} \int p^4 dp\, u_{12}^2 du_{12}\; H_4 = 
    \sum_{\ell} (-1)^{\ell} (2\ell + 1) \mathcal{L}_{\ell}(\hat{R} \cdot \hat{R}') \, \int \frac{p^2 dp}{2\pi^2}\;j_{\ell}(p R) j_{\ell}(p' R') = \delta_{\rm D}^{[3]} ({\bf R} + {\bf R'}),
\end{align}
where the last equality follows from equation (\ref{eqn:seventyeight}).

Now, using equation (\ref{eqn:true_4}) to replace $H_4$ with the expansion we have in four spherical Bessel functions, we find that
\begin{align}
    \delta_{\rm D}^{[3]} ({\bf R} + {\bf R'}) &= \frac{16}{\pi} \sum_{\ell_1 \ell_2 \ell_{12}} \sum_{\ell_3 \ell_4}  (-1)^{(\ell_1 + \ell_2 + \ell_3 + \ell_4)/2}\, \mathcal{D}^{\rm P} \sqrt{2\ell_{12} + 1} \nonumber\\
&\times \int p^4 dp\, u_{12}^2 du_{12}\;
j_{\ell_1}(pr_1) j_{\ell_2}(pr_2) j_{\ell_3} (p'r_3) j_{\ell_4} (p' r_4) j_{\ell_{12}} (p u_{12}) j_{\ell_{12}} (p' u_{12})
 \nonumber\\
&\qquad \times \begin{pmatrix}
\ell_1 & \ell_2 & \ell_{12}\\
0 & 0 & 0
\end{pmatrix}
 \begin{pmatrix}
\ell_{12} & \ell_3 & \ell_4\\
0 & 0 & 0
\end{pmatrix}
\mathcal{P}_{\ell_1 \ell_2 (\ell_{12}) \ell_3 {\ell_4}}(\hat{r}_1, \hat{r}_2, \hat{r}_3, \hat{r}_4).
\end{align}
We have used the remark below equation (\ref{eqn:true_4}), that the sum $\ell_1 + \ell_2 + \ell_3 + \ell_4$ is even, to remove that phase factor, relative to equation (\ref{eqn:true_4}).

Now, performing the $u_{12}$ integration we find
\begin{align}
    \delta_{\rm D}^{[3]} ({\bf R} + {\bf R'}) &= \frac{16}{\pi}
\sum_{\ell_1 \ell_2 } \sum_{\ell_3 \ell_4}  (-1)^{(\ell_1 + \ell_2 + \ell_3 + \ell_4)/2}\, \mathcal{D}^{\rm P} \int p^2 dp\;
j_{\ell_1}(pr_1) j_{\ell_2}(pr_2) j_{\ell_3} (p r_3) j_{\ell_4} (p r_4)
 \nonumber\\
&\qquad \times  \sum_{\ell_{12}}  \sqrt{2\ell_{12} + 1}  \begin{pmatrix}
\ell_1 & \ell_2 & \ell_{12}\\
0 & 0 & 0
\end{pmatrix}
 \begin{pmatrix}
\ell_{12} & \ell_3 & \ell_4\\
0 & 0 & 0
\end{pmatrix}
\mathcal{P}_{\ell_1 \ell_2 (\ell_{12}) \ell_3 {\ell_4}}(\hat{r}_1, \hat{r}_2, \hat{r}_3, \hat{r}_4).
\end{align}
$\mathcal{D}^{\rm P}$ is defined in the paragraph below equation (\ref{eqn:naive_4}).

It is again notable that the intermediate  $\ell_{12}$ here is simply summed over, with no effect on the spherical Bessel overlap integral on the right-hand side; this is because the Delta function is totally symmetric under interchange of its four arguments.  

\subsection{Delta Function for $N=5$}
With the two subsections treating the $N = 4$ case, we have already displayed all the important conceptual components involved in isotropic basis expansions of the Delta function. For completeness, we now present the results for $N = 5$. We first integrate the left-hand side of equation (\ref{eqn:lhs_N_5}) against $u_{123}^2 du_{123}$, obtaining a spherically-symmetric Delta function setting $p'' = p'$. We then integrate over $u_{12}^2 du_{12}$, which gives a spherically symmetric Delta function setting $p' = p$. Overall we find that 
\begin{align}
\int u_{12}^2 du_{12} \int u_{123}^2 du_{123}\; P_5 &= (4\pi)^7 \left(\frac{\pi}{2}\right)^2 \frac{1}{p^4}  \sum_{\rm all} (-1)^{\ell + \ell' + (\ell + \ell' + \ell'')/2} j_{\ell}(pR) j_{\ell'}(pR') j_{\ell''}(pR''') \nonumber\\
&\qquad \times Y_{\ell m}(\hat{R}) Y_{\ell' m'}(\hat{R}') Y_{\ell'' m''}(\hat{R}'') \mathcal{G}_{\ell' \ell \ell''}^{m' m m''}.
\label{eqn:N5_delta_interm}
\end{align}
We used the remark below equation (\ref{eqn:lhs_N_5}), that the sum $\ell + \ell' + \ell''$ is even, to simplify a power of $i$ into a power of $(-1)$ above. Now, we note that if the parity of $\ell + \ell'$ is equal to that of $(\ell + \ell' + \ell'')/2$, the phase above will vanish. If $\ell + \ell'$ is odd, then so is $\ell''$ (by the zero-projective quantum number 3-$j$ symbol) and hence $(\ell + \ell' + \ell'')/2$ is odd, so the phase vanishes. Now, if $\ell + \ell'$ is even, so is $\ell'$ (again by the zero-projective quantum number 3-$j$ symbol), and hence $(\ell + \ell' + \ell'')/2$ is even, so again the phase vanishes. With this simplification in hand, and summing over the projective quantum numbers in the spherical harmonics times the Gaunt integral, we can see that the above relation is simply equation (\ref{eqn:N3_delta}) (the relation for the $N=3$ delta function) prior to the integration of that latter over $p^2 dp/(2\pi^2)$ and also multiplied by $(4\pi)^5 (\pi/2)^2 p^{-4}$ and with the replacements $\hat{r}_1 \to \hat{R}$, $\hat{r}_2 \to \hat{R}'$, and $\hat{r}_3 \to \hat{R}''$. Hence, we see that integrating equation (\ref{eqn:N5_delta_interm}) against $p^6 dp/(2\pi^2)$ and dividing by $(4\pi)^5 (\pi/2)^2$ will give a Dirac Delta function $\delta_{\rm D}^{[3]}(\vec{R} + \vec{R}' + \vec{R}'')$.

Hence, we may write that 
\begin{align}
\delta_{\rm D}^{[3]}(\vec{R} + \vec{R}' + \vec{R}'') = \frac{1}{(4\pi)^5 (\pi/2)^2} \int \frac{p^6 dp}{2\pi^2} \int u_{12}^2 du_{12} \int u_{123}^2 du_{123}\; P_5.
\end{align}
Having thus shown that applying the operations outlined above to the generating function for the $N=5$ isotropic basis functions indeed yields the desired Delta function, we now need to apply the same operations to the right-hand side of equation (\ref{eqn:rhs_N5}) to obtain the Delta function's expansion in the isotropic basis. We see that the $u_{123}$ integration will produce a Dirac Delta function setting $p'' = p'$, and then the $u_{12}$ integration will set $p' = p$. So we have overall a factor of $p^{-4}$ to deal with. Integrating against $p^6 dp/(2\pi^2)$, we find
\begin{align}
&\delta_{\rm D}^{[3]}(\vec{R} + \vec{R}' + \vec{R}'')  = (4\pi)^4  \sum_{\rm all} (-1)^{(\ell_1 + \ell_2 + \ell_3 + \ell_4 + \ell_5)/2} \int \frac{p^2 dp}{2\pi^2}\; j_{\ell_1}(pr_1) j_{\ell_2}(pr_2) j_{\ell_3}(pr_3) j_{\ell_4} (pr_4) j_{\ell_5} (pr_5) \nonumber\\
&\qquad \times \begin{pmatrix}
\ell_1 & \ell_2 & \ell_{12}\\
0 & 0 & 0
\end{pmatrix}
\begin{pmatrix}
\ell_{12} & \ell_3 & \ell_{123}\\
0 & 0 & 0
\end{pmatrix}
\begin{pmatrix}
\ell_{123} & \ell_4 & \ell_5\\
0 & 0 & 0
\end{pmatrix}
\mathcal{D}^{\rm P} \sqrt{ 2\ell_{12} + 1} \sqrt{2 \ell_{123} + 1} \nonumber\\
&\qquad \times \mathcal{P}_{\ell_1 \ell_2 (\ell_{12}) \ell_3 (\ell_{123}) \ell_4 \ell_5}(\hat{r}_1, \hat{r}_2, \hat{r}_3, \hat{r}_4, \hat{r}_5).
\end{align}
We notice that the parity of $\ell_1 + \ell_2 + \ell_3 + \ell_4 + \ell_5$ is even, as can be shown by manipulating the rules implied by the zero-projective quantum number 3-$j$ symbols and canceling the parities of the intermediates.

\section{Connection to Overlap Integrals of Multiple Spherical Bessel Functions}
\label{sec:overlap}
The foregoing work with the Dirac Delta function already suggests that there is a connection between the isotropic basis functions and overlap integrals of multiple sBfs; here we explore this connection further. A complete discussion of previous works on overlap integrals of spherical Bessel functions is presented in \cite{meigs}; here we focus only on the results of immediate use in the present work.
\subsection{$N = 2$}
We consider integrating both sides of our fundamental relation for the $N = 2$ functions. We have
\begin{align}
\int dp\; j_0(pR) = \frac{\pi}{2} \frac{1}{R} = (4\pi)^{3/2} \sum_{\ell} \sqrt{2\ell + 1}\; \mathcal{P}_{\ell \ell}(\hat{r}_1, \hat{r}_2) \int dp\; j_{\ell}(pr_1) j_{\ell}(pr_2).
\end{align}
The first equality is simply from changing variables and using the result for the Sine integral.\footnote{\url{https://mathworld.wolfram.com/SineIntegral.html}} By comparing $(\pi/2)(1/R)$ to the usual expansion of a two-argument denominator in Legendre polynomials, we may conclude that the two-sBf overlap integral on the right-hand side is $(\pi/2) r_<^{\ell}/r_>^{\ell+1}$, where $r_<$ is the lesser of $r_1$ and $r_2$ and $r_>$ the greater. This recovers the standard result for this sBf integral (\textit{e.g.} \cite{pt_decoupling} equation C8 and references below).

\subsection{$N = 3$}
We now consider the same method applied to the $N = 3$ functions. We have
\begin{align}
\label{eqn:e-seven}
\int dp\;j_0(pR)  = \frac{\pi}{2}\frac{1}{R} &= (4\pi)^2 \sum_{\ell_1 \ell_2 \ell_3} 
\begin{pmatrix}
\ell_1 & \ell_2 & \ell_3\\
0 & 0 & 0
\end{pmatrix}
 \mathcal{C}_{\ell_1 \ell_2 \ell_3} (-1)^{(\ell_1 + \ell_2 + \ell_3)/2}\, \mathcal{P}_{\ell_1 \ell_2 \ell_3}(\hat{r}_1, \hat{r}_2, \hat{r}_3)\nonumber\\
&\qquad \times \int dp\; j_{\ell_1} (pr_1) j_{\ell_2} (pr_2) j_{\ell_3}(pr_3), 
\end{align}
where we again used the Sine integral to evaluate the left-hand side. The right-hand side above is simply from equation (\ref{eqn:sixteen}), and we recall that $\mathcal{C}_{\ell_1 \ell_2 \ell_3}$ is defined below equation (\ref{eqn:near_calC}).

We now integrate both sides against $\mathcal{P}^*_{\ell_1 \ell_2 \ell_3}$. By orthonormality of the $\mathcal{P}_{\Lambda}$, this reduces the right-hand side to simply the triple-sBf overlap integral (with a constant pre-factor). 

Now, to obtain an explicit result in terms of $r_1, r_2$ and $r_3$ we consider expanding the $1/R$ on the left-hand side. We have, recalling the definition of $R$ (equation \ref{eqn:R_Def} with $N = 3$), as $R \equiv |\vec{r}_1 + \vec{r}_2 + \vec{r}_3|$, that
\begin{align}
\label{eqn:e-eight}
\frac{1}{R} &= \frac{1}{|\vec{r}_1 + \vec{r}_2 +\vec{r}_3|} = I_0^0(\vec{r}_{12} + \vec{r}_3)  \nonumber\\
& =  \sum_{\lambda = 0}^{\infty} {{2\lambda + 1} \choose {2\lambda}} ^{1/2} \sum_{\mu = -\lambda}^{\lambda} R_{\lambda}^{\mu} (\vec{r}_{12}) I_{\lambda}^{-\mu} (\vec{r}_3) \left<\lambda, \mu; \lambda, -\mu| 0 0   \right>.
\end{align}
We have defined $\vec{r}_{12} \equiv \vec{r}_1 + \vec{r}_2$. $I_{\ell}^m$ is the irregular solid harmonic,
\begin{align}
    I_{\ell}^m(\vec{R}) &\equiv \sqrt{\frac{4\pi}{2\ell + 1}}\, \frac{Y_{\ell m}(\hat{R})}{R^{\ell + 1}};\\
    I_0^0 (\vec{R}) &= \sqrt{4\pi}\, \frac{Y_{00}}{R} = \frac{1}{R}.\nonumber
\end{align}
and $R_{\lambda}^{\mu}$ is the regular solid harmonic. On the second line of equation (\ref{eqn:e-eight}), the first factor within the sum is a binomial coefficient, and the final factor, in angle brackets, is a Clebsch-Gordan coefficient, standardly related to the 3-$j$ symbol\footnote{\textit{E.g.} \url{https://mathworld.wolfram.com/Clebsch-GordanCoefficient.html}} (but more convenient for our intermediate steps involving the shift theorem). For clarity, we use a semicolon to separate the orbital angular momentum-projective quantum number pair for the first and second of the two being added into the third, which appears after the bar.  We also duplicate here for convenience the relation between the Clebsch-Gordan coefficient and the 3-$j$ symbol:
\begin{align}
    \left<\ell_1, m_1; \ell_2, m_2 | \ell m \right> = (-1)^{m + \ell_1 - \ell_2} \sqrt{2\ell + 1} 
    \begin{pmatrix}
\ell_1 & \ell_2 & \ell\\
m_1 & m_2  & -m
\end{pmatrix}.
\end{align}

To obtain the second line of equation (\ref{eqn:e-eight}), we used the addition theorem for irregular solid harmonics. Two early proofs of this result are \citep{sack1964three} and \citep{weniger1985simple}; \citep{weniger2000addition} gives a treatment using Taylor series, while \citep{deb} uses the plane-wave expansion to prove the theorem. For related results on spherical expansions, we also note \citep{rowe1978spherical}. 

The result we have written requires that $r_{12} \leq r_3$. If $r_{12} > r_3$, one would interchange ${\bf r}_{12} \leftrightarrow {\bf r}_3$ in the second line. We now use the addition theorem for regular solid harmonics to expand $R_{\lambda}^{\mu}(\vec{r}_{12})$; this imposes no restriction on the magnitudes of $r_1$ and $r_2$.
We have 
\begin{align}
    \frac{1}{R} = \sum_{\lambda \mu} (-1)^{\lambda - \mu} I_{\lambda}^{- \mu}(\vec{r}_3) \sum_{\lambda' = 0}^{\lambda}  {2\lambda \choose 2\lambda'}^{1/2}
 \sum_{\mu' = -\lambda'}^{\lambda'} R_{\lambda'}^{\mu'}(\vec{r}_1)
 R_{\lambda - \lambda'}^{\mu - \mu'}(\vec{r}_2)
 \left<\lambda', \mu' ; \lambda - \lambda', \mu - \mu' | \lambda \mu \right>.
 \end{align}
We used the relationship between Clebsch-Gordan coefficients and 3-$j$ symbols,\footnote{\textit{E.g.} \url{https://mathworld.wolfram.com/Clebsch-GordanCoefficient.html} equation 7.} and then NIST DLMF 34.3.1, to compute the zero-total angular momentum Clebsch-Gordan coefficient, and we also evaluated the binomial coefficient that came from equation (\ref{eqn:e-eight}); all together these simplifications lead to the phase factor $(-1)^{\lambda -  \mu}$ above. 

Rewriting the solid harmonics explicitly using spherical harmonics and powers of the $r_i$, we find
\begin{align}
\label{eqn:one-over-sec}
\frac{1}{R} &=  4\pi \sum_{\lambda \mu, \lambda' \mu'} 
\mathcal{C}_{\lambda, \lambda', \lambda - \lambda'}^{-1}
{{2\lambda}\choose{2\lambda'}}^{1/2}
\begin{pmatrix}
\lambda' & \lambda - \lambda' & \lambda\\
\mu' & \mu - \mu' & -\mu
\end{pmatrix}
r_3^{-\lambda - 1}\, r_1^{\lambda'}\, r_2^{\lambda - \lambda'}\\
&\qquad \qquad \qquad \qquad \times Y_{\lambda, -\mu}(\hat{r}_3)
Y_{\lambda'\mu'}(\hat{r}_1)
Y_{\lambda - \lambda', \mu - \mu'}(\hat{r}_2).\nonumber
\end{align}
 We recall that $r_{12} \leq r_3$ here. We note that our original integral has no preference for $\ell_1$, $\ell_2$, or $\ell_3$.  

Now, following our prescription below equation (\ref{eqn:e-seven}), we integrate equation (\ref{eqn:one-over-sec}) against $\mathcal{P}^*_{\ell_1 \ell_2 \ell_3}$. This will force $\lambda', \mu' = \ell_1, m_1$, $\lambda - \lambda', \mu - \mu' = \ell_2, m_2$, and $\lambda, -\mu = \ell_3, m_3$ by orthogonality of the spherical harmonics, and thus eliminate the sums over $\lambda, \mu$ and $\lambda', \mu'$. It also imposes the constraint that $\ell_3 - \ell_1 = \ell_2$, by substituting for $\lambda$ and $\lambda'$ in the middle equality just stated. 

We may now sum over the projective quantum numbers using NIST DLMF 34.3.18, and find 
\begin{align}
\label{eqn:nine-two}
    \int d\hat{r}_1 d\hat{r}_2 d\hat{r}_3\;  \frac{1}{R}\;\mathcal{P}_{\ell_1 \ell_2 \ell_3}^*(\hat{r}_1, \hat{r}_2, \hat{r}_3) = 4\pi\, \mathcal{C}_{\ell_1 \ell_2 \ell_3}^{-1} {2\ell_3 \choose 2\ell_1}\, r_3^{-\ell_3 - 1} \,r_1^{\ell_1} \, r_2^{\ell_2} \;\delta^{\rm K}_{\ell_3, \ell_1 + \ell_2}.
\end{align}
We highlight that the integral result we have obtained above is simply the expansion coefficient of $1/R$ in the isotropic basis functions of three arguments. This conclusion, along with the form of the expansion, emphasizes the point that the isotropic basis functions are really just a generalization of the Legendre polynomials to more arguments. We notice that, analogously to the Legendre series, the larger length (here $r_3$) is taken to a power one larger in the denominator than the total power $(\ell_1 + \ell_2)$ of the smaller lengths in the numerator.

We also note that one could pursue the same steps if one had split the problem as $\vec{r}_{13} \equiv \vec{r}_1 + \vec{r}_3$ and $\vec{r}_2$, or even $\vec{r}_{23} \equiv \vec{r}_2 + \vec{r}_3$ and $\vec{r}_1$ and one would find equivalent answers. They will pertain respectively to regions on which $r_3 > r_{12}$, $r_2 > r_{13}$ and $r_1 > r_{23}$. Since there is full interchange symmetry among the vectors $\vec{r}_i$, we can simply choose an ordering and so will only need one result; we may therefore adopt the one explicitly obtained above. 

Setting equation (\ref{eqn:nine-two}) (times $\pi/2$) equal to equation (\ref{eqn:e-seven}) and solving for our desired integral overlap integral of the three sBfs, which we denote $I_{\ell_1 \ell_2 \ell_3}^{[0]}$, where the zero indicates the power of $p$ weighting the integrand, we find
\begin{align}
    I_{\ell_1 \ell_2 \ell_3}^{[0]}(r_1, r_2, r_3) &= 
    \frac{1}{8}\, \mathcal{C}_{\ell_1 \ell_2 \ell_3}^{-2} 
    \begin{pmatrix}
\ell_1  & \ell_2 & \ell_3\\
0 & 0 & 0
\end{pmatrix}^{-1}
{2\ell_3 \choose 2\ell_1}\,
(-1)^{(\ell_1 + \ell_2 + \ell_3)/2}\,
 r_3^{-\ell_3 - 1} \,r_1^{\ell_1} \, r_2^{\ell_2}\;\delta^{\rm K}_{\ell_3, \ell_1 + \ell_2},
 \\&\qquad\qquad\qquad\qquad\qquad\qquad\qquad\qquad
\qquad\qquad\qquad\qquad\qquad r_{12} < r_3,\nonumber
\end{align}
and where there is no constraint between $r_1$ and $r_2$. This result is analogous to that of \cite{mehrem_2_91}
but with a weight in the integral of $p^0$ rather than their case, which would have $p^2$ in our notation; discussion of this latter type of integral from a different perspective is also in \cite{jackson_maximon_72}.


\subsection{$N = 4$}
We focus on equation (\ref{eqn:true_4}). We first compute the integral of the left-hand side over $u_{12}^2 du_{12}$; this integration yields a factor of $\pi/(2 p p') \delta_{\rm D}^{[1]}(p - p')$. Integrating over $p'^2 dp'$ what remains on the left-hand side against this Dirac Delta function factor, we then find 
\begin{align}
\mathcal{J}_4 \equiv \int p'^2 dp'\;\int u_{12}^2 du_{12}\; H_4 = \frac{\pi}{2} \sum_{\ell} (2\ell + 1) \mathcal{L}_{\ell}(\hat{R} \cdot \hat{R}') j_{\ell}(pR) j_{\ell}(pR').
\end{align}
Finally, integrating over $dp$, we obtain 
\begin{align}
\bar{\mathcal{J}}_4 \equiv \int  dp\; \mathcal{J}_4 &= \frac{\pi}{2} \sum_{\ell} ( 2\ell + 1) \mathcal{L}_{\ell}(\hat{R} \cdot \hat{R}')  \int dp\;j_{\ell}(pR) j_{\ell}(pR') = \left(\frac{\pi}{2}\right)^2 \sum_{\ell} \mathcal{L}_{\ell}(\hat{R} \cdot \hat{R}')  \frac{R_<^{\ell}}{R_>^{\ell+1}}\nonumber\\
&=  \left( \frac{\pi}{2}\right)^2 \frac{1}{|\vec{r}_1 + \vec{r}_2  - (\vec{r}_3 + \vec{r}_4)|}.
\label{eqn:inv_4}
\end{align}
For the integral over $dp$, we used the result equation (C8) from \cite{pt_decoupling}.\footnote{The integral there has a typo; each $L_1$ inside the integral should be $L$.} We also used equation (\ref{eqn:multip}) to go from the Legendre series above to the final line above. We now observe that the fraction $1/|\vec{r}_1 + \vec{r}_2 - (\vec{r}_3 + \vec{r}_4)|$ is simply the irregular solid harmonic of order and degree zero, $I_0^0(\vec{R} - \vec{R}')$, with $\vec{R} \equiv \vec{r}_1 + \vec{r}_2$ and $\vec{R}' \equiv \vec{r}_3 + \vec{r}_4$. We may then apply to it the shift theorem for the irregular solid harmonic; with $R \leq R'$, we find
\begin{align}
\label{eqn:ninety-six}
\bar{\mathcal{J}}_4 =   \left( \frac{\pi}{2}\right)^2 \sum_{\lambda \mu} \mathcal{D}^+_{0,\lambda,0,\mu}\; R_{\lambda}^{\mu}(\hat{R}) I_{\lambda}^{-\mu}(\hat{R}')(-1)^{\lambda},
\end{align}
where the final phase stems from the sign on $\vec{R}'$.
We have defined 
\begin{align}
\mathcal{D}_{\lambda_1, \lambda_2, \mu_1, \mu_2}^+ &\equiv {{2(\lambda_1 + \lambda_2) + 1} \choose {2\lambda_2}}^{1/2} \left< \lambda_2, \mu_2; \lambda_1 + \lambda_2, \mu_1 - \mu_2 | \lambda_1 \mu_1 \right>,
\end{align}
as it will be of use in what follows; in equation (\ref{eqn:ninety-six}) we have the limit where $\lambda_1 = 0, \lambda_2 = \lambda, \mu_1 = 0, \mu_2 = \mu$.

Each solid harmonic on the right-hand side of equation (\ref{eqn:ninety-six}) can be expanded using the appropriate shift theorems (there is one for the regular solid harmonic, $R_{\lambda}^{\mu}$, and another for the irregular solid harmonic, of which we already once made use). We require $r_3 \leq r_4$ for this latter expansion. We find
\begin{align}
\bar{\mathcal{J}}_4 = & \left( \frac{\pi}{2}\right)^2 \sum_{\lambda \mu} \mathcal{D}_{0, \lambda, 0, \mu}^+ \sum_{\lambda' \mu'} \mathcal{D}_{\lambda ,\lambda', \mu ,\mu'} \sum_{\lambda'' \mu''} \mathcal{D}^+_{\lambda, \lambda'', -\mu, \mu''} \nonumber\\
&\times (4\pi)^2 \left[(2\lambda' + 1) (2 (\lambda - \lambda') + 1)(2\lambda'' + 1) (2(\lambda' + \lambda'') + 1)\right]^{1/2} r_1^{\lambda'} r_2^{\lambda - \lambda'} r_3^{\lambda''}r_4^{-\lambda - \lambda'' - 1}\nonumber\\
&\times Y_{\lambda' \mu'}(\hat{r}_1) Y_{\lambda - \lambda', \mu - \mu'}(\hat{r}_2) Y_{\lambda'', \mu''}(\hat{r}_3) Y_{\lambda + \lambda'', -\mu - \mu''}(\hat{r}_4)(-1)^{\lambda}.
\label{eqn:lhs_res_4}
\end{align}
The constants in the middle line stem from the normalizations of the four solid harmonics that were ultimately involved. We have also defined 
\begin{align}
\mathcal{D}_{\lambda_1, \lambda_2, \mu_1, \mu_2} &\equiv {{2\lambda_1} \choose {2\lambda_2}}^{1/2} \left< \lambda_2, \mu_2; \lambda_1 - \lambda_2, \mu_1 - \mu_2 | \lambda_1 \mu_1 \right>;
\end{align}
we note it is different from $\mathcal{D}^+$ as defined in equation (\ref{eqn:ninety-six}).

We now integrate the right-hand side of equation (\ref{eqn:true_4}) against $u_{12}^2 du_{12}$ and then against $p'^2 dp'$, just as we did for the left-hand side above. We then integrate over $dp$, finding
\begin{align}
\bar{\mathcal{J}}_4 &= (4\pi)^2 \left( \frac{\pi}{2} \right) \int  dp\; j_{\ell_1}(pr_1) j_{\ell_2}(pr_2) j_{\ell_3}(pr_3) j_{\ell_4}(p r_4) 
\mathcal{D}^{\rm P} \sqrt{2\ell_{12} + 1} \nonumber\\
&\qquad \times (-1)^{(\ell_1 + \ell_2 + \ell_3 + \ell_4)/2} \begin{pmatrix}
\ell_1 & \ell_2 & \ell_{12}\\
0 & 0 & 0
\end{pmatrix}
 \begin{pmatrix}
\ell_{12} & \ell_3 & \ell_4\\
0 & 0 & 0
\end{pmatrix}
 \mathcal{P}_{\ell_1 \ell_2 (\ell_{12}) \ell_3 \ell_4}(\hat{r}_1, \hat{r}_2, \hat{r}_3, \hat{r}_4).
 \label{eqn:rhs_inv_4}
\end{align}
Notably, comparing equations (\ref{eqn:inv_4}) and (\ref{eqn:rhs_inv_4}) reveals an expansion of the inverse of a sum of four vectors in terms of the $N = 4$ isotropic functions; this is another illustration that these latter really are generalizations of the Legendre polynomials.

Now, setting equations (\ref{eqn:rhs_inv_4}) and (\ref{eqn:lhs_res_4}) equal and then integrating both sides against $\mathcal{P}_{\ell_1 \ell_2 (\ell_{12}) \ell_3 \ell_4}^*$ sets $\lambda' = \ell_1,\; \lambda - \lambda' = \ell_2,\;\lambda'' = \ell_3$ and $\lambda + \lambda'' = \ell_4$, which, taken together, imply that $\ell_1 + \ell_2 + \ell_3 = \ell_4$. This gives an expression for the overlap integral of four spherical Bessel functions, weighted by $p$, in terms of a sum over powers of the free arguments $r_i$. Explicitly, we obtain
\begin{align}
    &\int  dp\; j_{\ell_1}(pr_1) j_{\ell_2}(pr_2) j_{\ell_3}(pr_3) j_{\ell_4}(p r_4) = \left[ \mathcal{D}^{\rm P} \sqrt{2 \ell_{12} + 1} \right]^{-1} \begin{pmatrix}
\ell_1 & \ell_2 & \ell_{12}\\
0 & 0 & 0
\end{pmatrix}^{-1}
 \begin{pmatrix}
\ell_{12} & \ell_3 & \ell_4\\
0 & 0 & 0
\end{pmatrix}^{-1}\nonumber\\
&\qquad \times \frac{\pi}{2} \sum_{\lambda \mu} \sum_{\lambda' \mu'} \sum_{\lambda'' \mu''} 
\mathcal{D}_{0, \lambda, 0, \mu}^+  \mathcal{D}_{\lambda ,\lambda', \mu ,\mu'} \mathcal{D}^+_{\lambda, \lambda'', -\mu, \mu''} (-1)^{\lambda}\nonumber\\
&\qquad \times \left[(2\lambda' + 1) (2 (\lambda - \lambda') + 1)(2\lambda'' + 1) (2(\lambda' + \lambda'') + 1)\right]^{1/2} r_1^{\lambda'} r_2^{\lambda - \lambda'} r_3^{\lambda''}r_4^{-\lambda - \lambda'' - 1} \nonumber\\
&\qquad \times \int d\Omega_R\; Y_{\lambda', \mu'}(\hat{r}_1) \cdots Y_{\lambda + \lambda'', -\mu - \mu''}(\hat{r}_4)\, \mathcal{P}^*_{\ell_1 \ell_2 (\ell_{12}) \ell_3 \ell_4}(\hat{r}_1, \hat{r}_2, \hat{r}_3, \hat{r}_4).
\label{eq:one-oh-one}
\end{align}
The last line may be evaluated using the definition of the $N=4$ isotropic basis function, and the Kronecker deltas resulting from the orthogonality of the spherical harmonics involved will eliminate the sums over $\lambda, \lambda', \lambda'', \mu, \mu'$ and $\mu''$. Importantly, there will then be no infinite sums.

We find that 
\begin{align}
\lambda' = \ell_1,\;\;\;\lambda - \lambda' = \ell_2,\;\;\;\lambda'' = \ell_3, \;\;\;\lambda + \lambda'' = \ell_4.     
\end{align}
We have two relations that involve $\lambda$ and we may use them both to find that
\begin{align}
    \lambda = \ell_1 + \ell_2 = \ell_4 - \ell_3, 
\end{align}
which implies that
\begin{align}
    \ell_4 = \ell_1 + \ell_2 + \ell_3.
\end{align}
For completeness, we also note that the angular integration in Eq. (\ref{eq:one-oh-one}) leads to
\begin{align}
    \mu  = m_1 + m_2, \;\;\; \mu' = m_1,\;\;\; \mu'' = m_3.
\end{align}

Now, the 4-argument isotropic basis function Eq. (\ref{eq:one-oh-one}) includes two 3-$j$ symbols and, given the relation on the angular  momenta above, the triangle inequalities on those  symbols can only be satisfied if the intermediate is $\ell_{12} =  \ell_1 + \ell_2$. With this in hand, we see that both 3-$j$ symbols from the basis function must be ``flattened'' (with upper rows, respectively, $\ell_1, \ell_2, \ell_1 + \ell_2$ and $\ell_1 + \ell_2, \ell_3, \ell_1 + \ell_2 + \ell_3$). It turns out they will match the two 3-$j$ symbols that come from the middle and last $\mathcal{D}$ coefficients. 

We also notice that the first $\mathcal{D}$ does not depend on any projective quantum numbers. Hence we may sum over the projective quantum numbers for the four 3-$j$ symbols (two from the $\mathcal{D}$, two  from  the isotropic basis function) that \textit{do} depend on them. Since we have two matching pairs of these symbols, as noted above, this sum just gives unity by NIST DLMF 34.3.18.

We are then left with the first $\mathcal{D}$ only, and after substituting our values for $\lambda$ and $\mu$ and simplifying it becomes
\begin{align}
    \mathcal{D}_{0, \lambda, 0, \mu}^+ = (-1)^{\ell_1 + \ell_2}\sqrt{\frac{2(\ell_1 + \ell_2)}{2(\ell_1 + \ell_2)+1}}. 
    \label{eq:simp_D}
\end{align}

Using this result,  recalling the definition of $\mathcal{D}^{\rm P}$ (which will cancel with the pre-factors  in the third line of equation \ref{eq:one-oh-one}), and simplifying, we find
\begin{align}
    &\int  dp\; j_{\ell_1}(pr_1) j_{\ell_2}(pr_2) j_{\ell_3}(pr_3) j_{\ell_4}(p r_4) =  \frac{\pi}{2} 
 \begin{pmatrix}
\ell_1 & \ell_2 & \ell_{1} + \ell_2\\
0 & 0 & 0
\end{pmatrix}^{-1}
 \begin{pmatrix}
\ell_{1} +  \ell_2 & \ell_3 & \ell_1 + \ell_2 + \ell_3\\
0 & 0 & 0
\end{pmatrix}^{-1}\nonumber\\
&\qquad\qquad\times (-1)^{\ell_1 + \ell_2}
\sqrt{2(\ell_1 + \ell_2)} \sqrt {2(\ell_1 + \ell_2)+1} {2(\ell_1 + \ell_2) \choose 2\ell_1 }^{1/2}
{2(\ell_1 + \ell_2 + \ell_3) \choose 2\ell_3}^{1/2}
\nonumber\\
&\qquad\qquad\times r_1^{\ell_1} r_2^{\ell_2} r_3^{\ell_3}
r_4^{-\ell_1 - \ell_2  - \ell_3  -1}. 
\end{align}
The definition of the 4-argument isotropic function contained factors that cancelled the phase in Eq. (\ref{eq:simp_D}) and also the factor of $[\sqrt{2\ell_{12} + 1}]^{-1}$ in Eq. (\ref{eq:one-oh-one}). To avoid having the 3-$j$ symbols vanish  (and thus their inverses, diverge),  $\ell_1$  and $\ell_2$ must have the same parity,  and the same goes for $\ell_1 + \ell_2$ and $\ell_3$. Hence we see that only integrals where  the sum $\ell_1 + \ell_2  + \ell_3 + \ell_4$ is even give sensible results by this method; however, since we found that $\ell_4 = \ell_1 + \ell_2 + \ell_3$, this condition will also be satisfied.\footnote{NIST DLMF 34.3.6 gives a formula in terms of factorials for the flattened  3-$j$ symbols  in the first line, and  while a few cancellations will occur with the factorials from the binomial coefficients in the second line, we did not regard the resulting expression as notably simpler.} Finally, we note that a related integral, over four sBfs with a $p^2$ weight, was evaluated in \cite{mehrem_4_pub} as a finite sum over an associated Legendre function.

\section{Use in Enabling a Faster NPCF Algorithm for $N \geq 4$}
\label{sec:alg}
Recently, an algorithm to compute N-Point Correlation Functions (NPCFs), $N\geq 4$ has been proposed \cite{encore}, based on the spherical-harmonic 3PCF algorithm of \cite{3pt_alg, aniso_3, 3pcf_fft, portillo} (which was already applied to observational data in \cite{se_3pcf_boss, se_3pcf_bao, se_rv_boss} using models in \cite{se_rv, se_3pcf_rsd}). This algorithm has already been used to make the first measurement of the connected 4PCF \cite{4pcf_msmt}. This algorithm projects the multi-point correlation functions onto the isotropic basis functions of one less index; \textit{i.e.} an N-Point Correlation Function (NPCF) would require the $(N-1)$-argument basis functions. This is because one point can be placed at the origin of coordinates. The algorithm proceeds by sitting on each galaxy in a survey, one at a time (the ``primary"), and then taking its neighbors out to some maximum distance $R_{\rm max}$ away and placing them in spherical shells (``bins"). On each shell, the angular dependence of the pattern (about the primary) of galaxies on that shell is expanded into spherical harmonics, with coefficients $a_{\ell m}$. These coefficients may then be combined, weighted by a 3-$j$ symbol or an appropriate combination of 3-$j$ symbols, to form the projection of the NPCF about that primary onto the isotropic basis. This process is repeated on every primary and the results are then averaged.

This algorithm is quite fast because it never explicitly needs to consider $(N-1)$-tuplets about a primary; obtaining the harmonic expansion coefficients about each primary scales as $nV_{\rm max}$ where $n$ is the number density and $V_{\rm max}$ is the spherical volume out to $R_{\rm max}$ about the primary, and must be done per primary, so the total scaling is $N_{\rm g} nV_{\rm max}$ if there are $N_{\rm g}$ galaxies in the survey. However, about each primary, the appropriately-weighted combination of $(N-1)$ coefficients $a_{\ell m}$ must be formed, and this must be done for all bin combinations. Put more abstractly, if we consider the $a_{\ell m}$ at each fixed $\ell$ and $m$ a vector of length $N_{\rm bins}$, then we form the outer product on bins of the $a_{\ell m}$. Furthermore, at each fixed combination of bins, there are many different combinations of $\ell$ that are permitted by isotropy. And the combinations of $a_{\ell m}$ must be summed over projective quantum numbers (this comes from the fact that the clustering should be agnostic as to the $z$-axis, yet the projective quantum numbers know about the $z$-axis---so they must be summed out).

Overall, this forming many $\ell$ combinations and then summing over projective quantum numbers leads to a complicated, multi-level loop. In practice, in the \textsc{encore} code, one has a nested 9-level loop. Fundamentally there are selection rules on the $\ell$ combinations, and so one cannot unroll and parallelize the loops any further. Consequently on the CPU, this step takes the dominant amount of time for 4PCF and higher.

Here, we suggest an algorithm that is less efficient, judged in terms of FLOPS, but may be faster on the CPU, and takes advantage of the generating functions presented above. If each isotropic basis function could be written directly in Cartesian coordinates, as the generating functions indeed make straightforward, then the orthogonality integral used to project the full NPCF onto the basis could be done directly at each combination of radial bins, with no need for any loops over the $\ell$ or the $m$. This approach would be less efficient in terms of FLOPS, because the Cartesian elements from which basis function is built are not orthogonal to those for other basis functions. Thus, one would repeat computation. The advantage of the spherical-harmonic-based approach is that one never repeats any computation---each $a_{\ell m}$ is independent and computed once. However, computation is not the limiting step in the 4PCF and higher algorithm, so there is presumably room to repeat some of it to enable avoiding the slow nested loops. 

However, there is still the problem of factorization: naively, the Cartesian isotropic basis functions involve dot and cross products of pairs of unit vectors, and so might require forming an $(N-1)$-tuplet about a primary to be able to perform the projection integral; this would scale catastrophically. Here, however, we may use the trick originally developed by \cite{Bianchi}, where one multiplies out the components of the Cartesian dot and cross products, and thus has a sum of integrals, each of which can be factorized. For instance, $\mathcal{P}_{112}$ involves $(\hat{r}_1 \cdot \hat{r}_3) (\hat{r}_2 \cdot \hat{r}_3)$ (see \cite{Cahn}, Appendix A, equation A.2). Multiplying this out, we find 
\begin{align}
\mathcal{P}_{112} \supset (\hat{r}_1 \cdot \hat{r}_3) (\hat{r}_2 \cdot \hat{r}_3) = r_{1x}^2 r_{2x} r_{3x} + r_{1x} r_{1y} r_{2y} r_{3x} + \cdots + r_{1z}^2 r_{2z} r_{3z};
\end{align}
$r_{ix}$ is the $x$-component of the vector $\vec{r}_i$, \textit{etc}. 

If we then wish to integrate the product of binned densities, $\bar{\delta}(R_i; \hat{r}_i)$ (where the bin is designated by $R_i$ and the bar denotes ``binned''), against $\mathcal{P}^*_{112}$, we may form integrals like 
\begin{align}
\int d\Omega_1\;\bar{\delta}(R_1; \hat{r}_1) r_{1x}^2 \int d\Omega_2 \;\bar{\delta}(R_2; \hat{r}_2) r_{2x} \int d\Omega_3 \;\bar{\delta}(R_3; \hat{r}_3) r_{3x} + \cdots.
\end{align}
For a discrete density field such as galaxies, these integrals simply become sums over each galaxy in the shell, weighted by the appropriate power of $r_{ix}$, \textit{etc}. Since they are factorized, they allow one to avoid examining an $(N-1)$-tuplet around each primary. Indeed, obtaining each integral simply scales as $n V_{\rm max}$ just as in the spherical-harmonic-based algorithm. 

Now, in forthcoming work (Slepian, Warner, Hou \& Cahn in prep.), we outline a GPU-based implementation of \textsc{encore} called \textsc{cadenza}. One might wonder if the above scheme would offer improvements for \textsc{cadenza} in the same way we suspect it will for \textsc{encore}. However, the GPU code never loops over the $\ell$ and $m$. GPUs are able to run a large number of concurrent threads, of the order of 1.4 million for typical hardware (NVIDIA A100), so each combination of the $\ell$ and $m$ is unrolled to a separate thread and already computed in parallel on the GPU. Thus, there is no further advantage to be gained using the scheme outlined above, whose primary goal was to avoid the loops used on the \textit{CPU}. Finally, we comment that another way around the challenge of the selection rules that the 4PCF here presents is to pursue instead a compressed statistic that does not require any angular momenta, for instance as recently presented in \cite{pop}.

\section{Concluding Discussion}
\label{sec:concs}
The isotropic basis functions \cite{Cahn} offer a natural basis for expanding functions of $N$ unit vectors that are invariant under joint rotations of all the arguments. 
Here, we have shown systematically how to generate isotropic basis functions' representations as sums of simple dot products (for the even-parity sector) by expanding a plane wave in two different ways, and then taking Taylor series and matching power by power. For the odd-parity sector (possible for basis functions with three arguments or more), the presence of a scalar triple product complicates this. However we have shown (\S\ref{sec:odd}) that given the lowest-lying parity-odd basis function, one can solve for the other parity-odd ones given the even ones (as obtained from our generating function procedure) and the linearization identity for products of isotropic basis functions (in \cite{Cahn}). 

The introduction discusses a number of applications that have already been made of the isotropic basis functions, specifically in measuring galaxy correlations. The main utility of the present work is that it enables easily obtaining simple, dot product (``Cartesian'') expressions for the isotropic basis functions. These expressions are not easily obtained from the spherical-harmonic-based definition of the isotropic basis functions presented in \cite{Cahn}. In turn, the Cartesian expressions can enable a speedier CPU-based algorithm for measuring correlations in the basis, which avoids the many-layered, highly-restricted loop required by the selection rules of the 3-$j$ symbols involved in the current approach to computing galaxy clustering in the isotropic basis (\S\ref{sec:alg}). 

We have also given useful expansions of both multi-argument Dirac delta functions (\S\ref{sec:dirac}), and inverse powers of vector sums (\S\ref{sec:overlap}), in the isotropic basis, and shown the intimate link between these latter and spherical Bessel overlap integrals. We expect these expressions will find use in inflationary calculations of multi-point correlations, as well as cosmological perturbation theory for the late-time evolution of galaxy clustering (\textit{e.g.} the type of integrals discussed in \cite{pt_decoupling})

 \section*{Acknowledgments}
We thank RN Cahn for many useful conversations, and the Slepian research group members for helpful discussions over the past years. JH acknowledges funding from the European Union's Horizon 2020 research and innovation programme under the Marie Skłodowska-Curie grant agreement no.\ 101007633. AG acknowledges funding from NASA grant number 80NSSC24M0021.

\bibliographystyle{unsrt}
\bibliography{refs}

\end{document}